\documentclass{article}
\usepackage{iclr2027_conference,times}
\usepackage[T1]{fontenc}
\usepackage[utf8]{inputenc}
\usepackage{amsmath,amssymb,amsthm,bm}
\usepackage{graphicx,booktabs,tabularx,multirow,array}
\usepackage[table]{xcolor}
\usepackage{hyperref,url}
\usepackage{microtype}
\usepackage{placeins,longtable}
\hypersetup{hidelinks,pdftitle={SPA: Structure-Preserving Approximation of Gated Activations under FHE},pdfauthor={Rui Li; Wenyuan Wu; Weijie Miao}}
\newtheorem{theorem}{Theorem}

\newcommand{\yes}{\ensuremath{\bullet}}
\newcommand{\no}{\ensuremath{\circ}}
\newcommand{\unk}{\textemdash}
\newcommand{\SPA}{\textsc{SPA}}
\newcommand{\erf}{\operatorname{erf}}

\definecolor{secblue}{RGB}{232,242,255}
\newcolumntype{L}[1]{>{\raggedright\arraybackslash}p{#1}}
\newcolumntype{Y}{>{\raggedright\arraybackslash}X}
\newcommand{\CodeRepositoryURL}{https://github.com/Anonymous000papers/Code-For-SPA}
\newcommand{\CodeRepositoryLink}{%
  \expandafter\url\expandafter{\CodeRepositoryURL}}

\newcommand{\PaperTitle}{SPA: Structure-Preserving Approximation of Gated Activations under FHE}
\title{\PaperTitle}
\newcommand{\PaperAuthors}{%
\begin{minipage}[t]{\dimexpr\textwidth-2\tabcolsep\relax}
\normalfont\raggedright
{\bfseries Rui Li\textsuperscript{1,2}\qquad
Wenyuan Wu\textsuperscript{1}\qquad
Weijie Miao\textsuperscript{3}}\par
\vspace{6pt}
{\small
\textsuperscript{1}Chongqing Key Laboratory of Secure Computing for Biology,\\
Chongqing Institute of Green and Intelligent Technology,\\
Chinese Academy of Sciences\par
\vspace{4pt}
\textsuperscript{2}Chongqing School, University of Chinese Academy of Sciences\par
\vspace{4pt}
\textsuperscript{3}Department of Industrial and Systems Engineering,\\
The Hong Kong Polytechnic University\par
\vspace{6pt}

}
\end{minipage}%
}
\author{\PaperAuthors}

\iclrfinalcopy
\begin{document}
\maketitle
\fancyhead[L]{Preprint}
\begin{abstract}
Nonlinear activation evaluation remains a major bottleneck in neural inference under fully homomorphic encryption (FHE). FHE supports efficient addition and multiplication, but lacks native operations for exponential functions, reciprocals, and value-dependent selection. High-degree polynomial approximations on fixed intervals are therefore commonly used to replace nonlinear activations. They can achieve small errors but often incur substantial multiplicative depth and latency. We observe that several widely used nonlinear activations share the algebraic structure $A(x)=xg(x)$ with $g(-x)=1-g(x)$. This class includes GELU, SiLU, fixed-parameter Swish, and the scalar nonlinear branches of GEGLU/SwiGLU. We propose \SPA{} (\emph{Structure-Preserving Approximation}) to exploit this shared structure. \SPA{} retains the exact odd component $x/2$ and approximates only the even residual as a polynomial in $x^2$, yielding a comparison-free arithmetic kernel. This construction removes redundant odd degrees of freedom and enables lower-cost polynomial evaluation at a given error target. 
Plaintext experiments with Gemma-4-E2B and Llama-3.2-3B show that SPA largely preserves downstream task performance without weight updates. Fixed-interval CKKS benchmarks show favorable error--latency trade-offs. On $[-5,5]$, \SPA{}-balanced achieves a $3.25\times$ speedup over MOAI (ICLR'~26) for erf-GELU with comparable error. On the same interval, \SPA{}-accurate achieves $79.22\%$ lower root mean squared error (RMSE) and a $10.40\times$ speedup over Euston (S\&P'~26) for QuickGELU. For SiLU, \SPA{}-accurate achieves $19.02\%$ lower RMSE and a $2.06\times$ speedup over THOR (CCS'~25).
\end{abstract}

\section{Introduction}
\label{sec:intro}
Fully homomorphic encryption (FHE) enables neural inference on encrypted inputs, but does not directly implement nonlinear activations such as GELU and SiLU. Their evaluation is a major challenge in large language model (LLM) feed-forward blocks \citep{ckks,orion,slothe}. The goal is to find accurate, low-cost replacements that balance polynomial accuracy and depth against lookup, selection, or communication costs \citep{iron,bolt,sigma,shaft}.

\begin{figure*}[ht]
\centering
\includegraphics[width=1.00\textwidth]{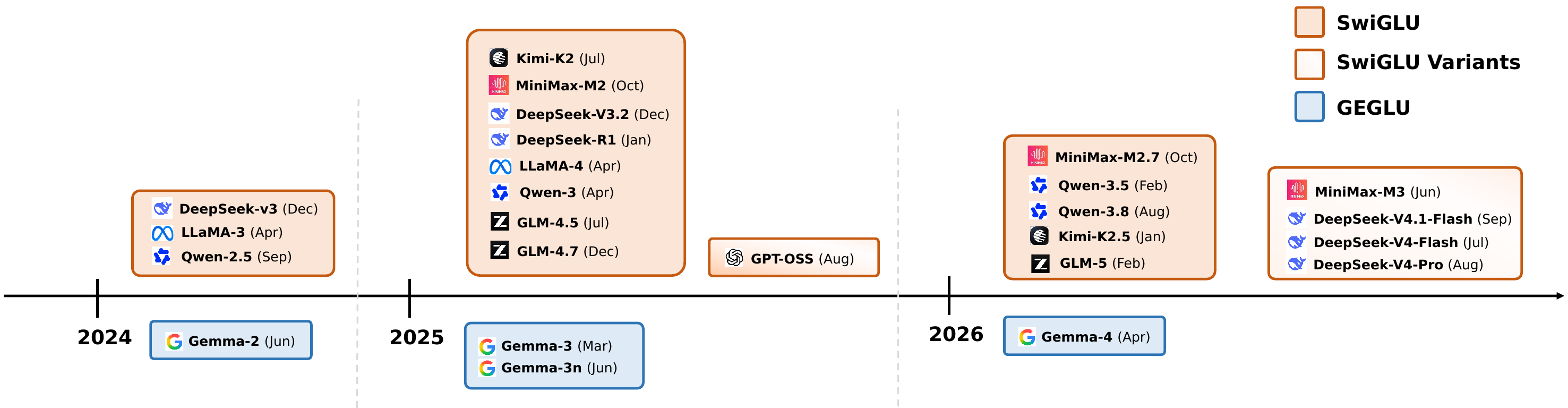}
\caption{Activation functions employed in several mainstream large language models. For multimodal models that may adopt different activation functions across modules, only those used in the text modules are reported. GEGLU/SwiGLU contain scalar GELU/SiLU branches, respectively.}
\label{fig:landscape}
\end{figure*}

We study the scalar GELU/SiLU branches in Figure~\ref{fig:landscape} and other activations with the shared structure
\begin{equation}
 A(x)=xg(x),\qquad g(-x)=1-g(x),
 \label{eq:class}
\end{equation}
where $g$ is a scalar \emph{gate function} that modulates $x$, not an additional learned projection. GELU, SiLU, and fixed-parameter Swish have this structure, which implies $A(x)-A(-x)=x$: the odd part is known exactly, and only the even remainder requires approximation. Replacing this scalar branch leaves the other projection, elementwise product, and trained weights unchanged.

We propose \SPA{}, \emph{Structure-Preserving Approximation}, which retains the linear odd part and approximates the even remainder after the change of variable $u=x^2$. This structural restriction preserves the best uniform approximation error at the corresponding degree in $x$, independently of the basis. Our implementation uses Chebyshev polynomials and discrete minimax fitting, with an explicit evaluation schedule and coefficient-specific error bounds.

\begin{samepage}
Our contributions are (i) a structure-preserving replacement based on~\eqref{eq:class}; (ii) best-approximation equivalence and finite-interval bounds for specified coefficients; and (iii) target-matched GELU, QuickGELU, and SiLU experiments assessing model performance and homomorphic error--cost trade-offs. An equivalent even-polynomial representation has the same approximation space.

\end{samepage}

\section{Related work}
\label{sec:related}
Table~\ref{tab:survey} summarizes representative methods. Degree, interval, and evaluation schedule affect homomorphic cost \citep{orion,slothe,thor,moai,euston}; range calibration reduces Sylph's SiLU approximation degree from 83 to 31 \citep{sylph}. Lookup and piecewise methods incur access or selection costs; MPC and hybrid protocols also require communication \citep{iron,bolt,sigma,nexus,mpcformer,shaft}.

\begin{table*}[ht]
\centering
\caption{Mechanisms in secure-activation methods. \yes: used; \no: not used in the described method; \unk: unverified. Entries summarize methods rather than rank complete systems.}
\label{tab:survey}
\small\setlength{\tabcolsep}{3.6pt}
\begin{tabular}{@{}ll*{7}{c}@{}}
\toprule
Work & Source & HE & MPC/FSS & GELU family & SiLU & Polynomial & Lookup & Selection \\
\midrule
Iron & NeurIPS 2022 & \yes & \yes & \yes & \unk & \no & \yes & \yes \\
MPCFormer & ICLR 2023 & \no & \yes & \yes & \unk & \yes & \no & \no \\
BOLT & IEEE S\&P 2024 & \yes & \yes & \yes & \unk & \yes & \no & \yes \\
SIGMA & PoPETs 2024 & \no & \yes & \yes & \yes & \no & \yes & \yes \\
SHAFT & NDSS 2025 & \no & \yes & \yes & \yes & \no & \no & \yes \\
NEXUS & NDSS 2025 & \yes & \no & \yes & \unk & \yes & \no & \yes \\
Orion & ASPLOS 2025 & \yes & \no & \unk & \yes & \yes & \no & \no \\
ENSI & arXiv 2025 & \yes & \no & \unk & \yes & \yes & \no & \no \\
THOR & CCS 2025 & \yes & \no & \yes & \unk & \yes & \no & \no \\
Sylph & arXiv 2026 & \yes & \no & \unk & \yes & \yes & \no & \no \\
MOAI & ICLR 2026 & \yes & \no & \yes & \yes & \yes & \no & \no \\
Euston & IEEE S\&P 2026 & \yes & \no & \yes & \unk & \yes & \no & \yes \\
\midrule
\SPA{} & This work & \yes & \no & \yes & \yes & \yes & \no & \no \\
\bottomrule
\end{tabular}
\end{table*}

Prior methods already exploit symmetry and residuals \citep{bolt,sigma,shaft}. The MOAI implementation approximates the factor $2g-1$ and reconstructs $x[1+\widehat f(x)]/2$; \SPA{} directly approximates the even residual $x[g(x)-1/2]$ as $q(x^2)$. An odd $\widehat f$ also preserves the exact odd part. The methods differ in their approximation targets and evaluation schedules.

Polynomial Transformer develops an HE-oriented architecture and operator-conversion procedure \citep{polytransformer}. Powerformer combines model distillation, pseudo-sign GELU/tanh approximations, and homomorphic matrix multiplication \citep{powerformer}. 
Their system timings cannot be compared directly with scalar-kernel timings. Powerformer’s GELU remains an untested baseline. OLA and ATLAS address the complementary problem of model-level allocation \citep{ola,atlas}.

\section{Structure-preserving polynomial approximation}
\label{sec:method}
\subsection{Activation structure and change of variable}
Fix $R>0$. Let $A(x)=xg(x)$ have the structure in~\eqref{eq:class} on $[-R,R]$, with $g\in C([-R,R])$. In particular, $g(0)=1/2$ and $A(0)=0$. Since $A(-x)=A(x)-x$, the odd and even parts of $A$ are, respectively,
\begin{equation}
 \frac{A(x)-A(-x)}{2}=\frac{x}{2},\qquad
 r_A(x):=\frac{A(x)+A(-x)}{2}=A(x)-\frac{x}{2}.
 \label{eq:evenodd}
\end{equation}
Since $r_A$ is even, it depends on $x$ only through $x^2$. With $u=x^2$, define
\begin{equation}
 H_A(u):=r_A(\sqrt{u})
       =\sqrt{u}\left[g(\sqrt{u})-\frac12\right],
 \qquad 0\le u\le R^2.
 \label{eq:residualu}
\end{equation}
The function $H_A$ is continuous, with $H_A(0)=0$. Since $r_A$ is even, $H_A(x^2)=r_A(|x|)=r_A(x)$, and hence
\begin{equation}
 A(x)=\frac{x}{2}+H_A(x^2),\qquad |x|\le R.
 \label{eq:decomp}
\end{equation}
The two functions describe the same residual. 
If $g$ is analytic, then $H_A$ is analytic near zero; continuity alone does not ensure this property (Appendix~\ref{app:analytic}).

We approximate $H_A$ by a polynomial $q$ and retain the linear term:
\begin{equation}
 \widehat A_q(x)=\frac{x}{2}+q(x^2).
 \label{eq:structured}
\end{equation}
Here, \emph{structure-preserving} refers to the exact odd part. The basis (e.g., Chebyshev) and objective (e.g., least squares or minimax) remain independent choices.

Our implementation maps $u\in[0,R^2]$ to $t\in[-1,1]$ by $t=2u/R^2-1$ and writes
\begin{equation}
 Q_c(t)=\sum_{j=0}^{k}c_jT_j(t),\qquad
 P_c(x)=\frac{x}{2}+Q_c\!\left(\frac{2x^2}{R^2}-1\right).
 \label{eq:profile}
\end{equation}
Here $T_j$ is the first-kind Chebyshev polynomial and $c_0$ is not halved. Degrees satisfy $\deg Q_c\le k$ and $\deg P_c\le2k$ for $k\ge1$. The representation reduces the number of free coefficients, not the degree in the original variable. Write $t_R(x)=2x^2/R^2-1$.

Let $\Pi_d$ denote the space of real polynomials of degree at most $d$, with $\|f\|_{\infty,I}=\max_{x\in I}|f(x)|$ on the stated compact intervals. Symmetry shows that~\eqref{eq:structured} preserves the best uniform approximation error \citep{trefethen}.
\begin{theorem}[Best-approximation equivalence and structural identities]
\label{prop:parity}
Let $A(x)=xg(x)$ have the structure in~\eqref{eq:class}, with $g\in C([-R,R])$, $R>0$, and $g(-x)=1-g(x)$. Define $H_A$ by~\eqref{eq:residualu}. For every integer $k\ge1$,
\begin{equation}
 \inf_{p\in\Pi_{2k+1}}\|A-p\|_{\infty,[-R,R]}
 =\inf_{p\in\Pi_{2k}}\|A-p\|_{\infty,[-R,R]}
 =\inf_{q\in\Pi_k}\|H_A-q\|_{\infty,[0,R^2]}.
 \label{eq:parity}
\end{equation}
Moreover, every polynomial $P_c$ in~\eqref{eq:profile} satisfies
\begin{equation}
 P_c(x)-P_c(-x)=x,\qquad P_c'(0)=\frac12.
 \label{eq:invariants}
\end{equation}
These identities remain valid after any change to the residual coefficients, including rounding, provided the linear term is unchanged; they are identities in real arithmetic.
\end{theorem}
Appendix~\ref{app:proofs} proves the result by symmetrization. The theorem characterizes the error attainable in these polynomial spaces; it does not guarantee that numerical fitting attains it.

\begin{figure*}[ht]
\centering
\includegraphics[width=0.90\textwidth]{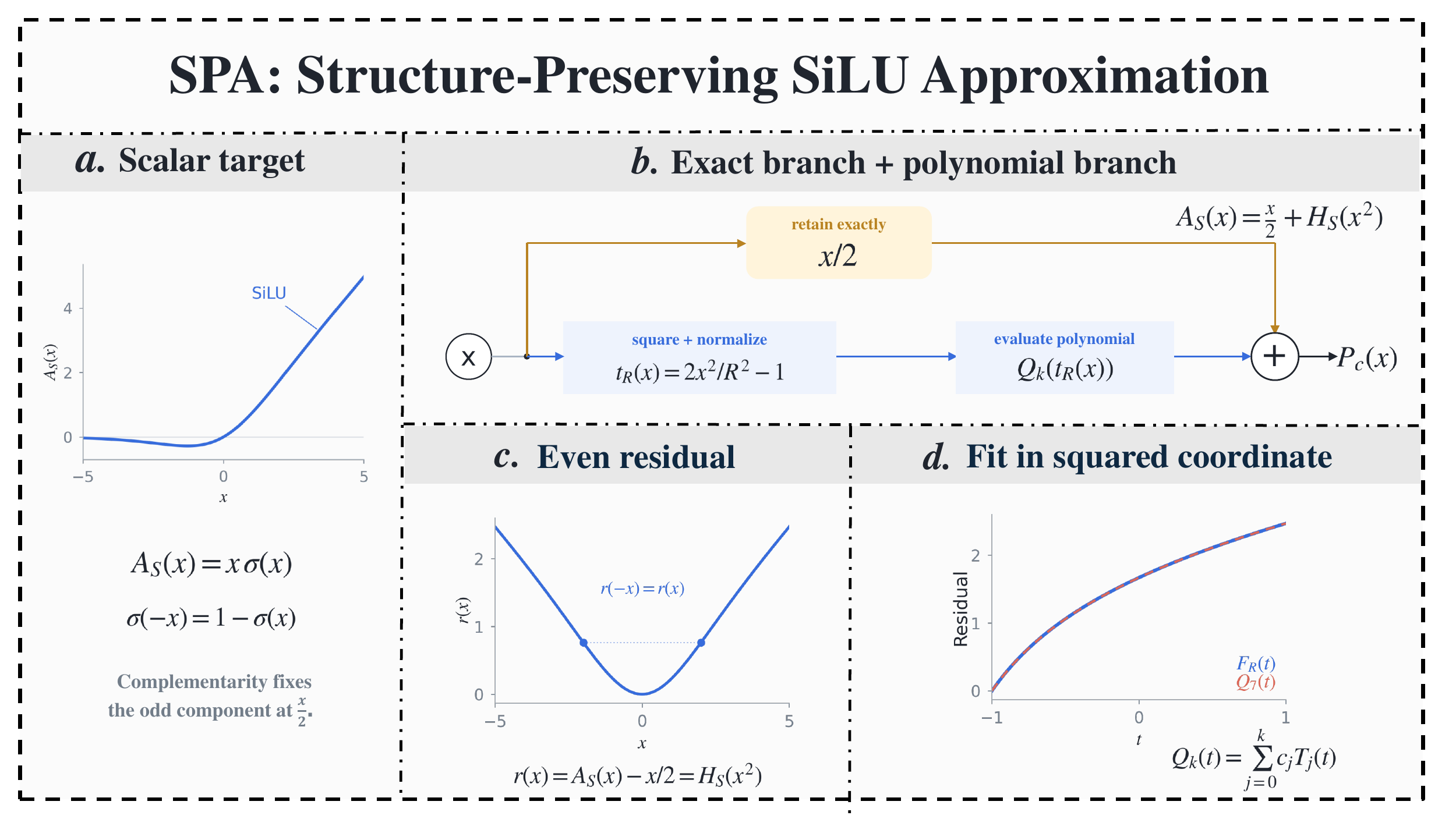}
\caption{Structure-preserving approximation, illustrated with SiLU. The linear term $x/2$ is retained; the even residual $r_A(x)=H_A(x^2)$ is approximated after the change of variable $u=x^2$. Chebyshev polynomials are the basis used in our implementation, not an assumption of the structural reduction.}
\label{fig:overview}
\end{figure*}

\subsection{Targets and coefficient construction}
\label{sec:fit}
The main experiments use three distinct target functions:
\begin{align}
 A_{\rm S}(x)&=x\sigma(x),&
 A_{\rm E}(x)&=\frac{x}{2}\bigl[1+\erf(x/\sqrt2)\bigr],\nonumber\\
 A_{\rm Q}(x)&=x\sigma(\beta x),&
 \sigma(x)&=(1+e^{-x})^{-1},\quad\beta=1.702.
 \label{eq:targets}
\end{align}
These are SiLU, erf-GELU, and QuickGELU, respectively.

For $F_R(t)=H_A(R^2(t+1)/2)$ and a finite set of nodes $G\subset[-1,1]$, the recorded implementation solves
\begin{equation}
 \min_{c\in\mathbb R^{k+1},\,\epsilon\ge0}\epsilon
 \quad\text{subject to}\quad
 |Q_c(t_i)-F_R(t_i)|\le\epsilon\quad(t_i\in G).
 \label{eq:discretefit}
\end{equation}
We solve this unweighted discrete minimax program with SciPy/HiGHS in float64. We fix split $m=4$ for $R=5$ or $m=8$ for $R\ge20$, increase block count $B=1,2,\ldots$, and test $k=Bm-1$ (Section~\ref{sec:psconstruction}). The first degree meeting a separate grid tolerance is retained: $10^{-2}$, $10^{-3}$, or $10^{-4}$ for fast, balanced, or accurate. These engineering tolerances are not model-accuracy thresholds.

\subsection{Homomorphic polynomial evaluation}
\label{sec:psconstruction}
For the chosen integer split $m\ge2$, repeated polynomial division defines blocks $B_b$ by
\begin{equation}
 Q_0=Q_c,\qquad Q_b=T_mQ_{b+1}+B_b,\quad\deg B_b<m,
 \label{eq:chebdivision}
\end{equation}
until $Q_B=0$, with $B=\lceil(k+1)/m\rceil$ and zero blocks padded if necessary. Substitution gives the Paterson--Stockmeyer representation \citep{ps}
\begin{equation}
 Q_c(t)=\sum_{b=0}^{B-1}B_b(t)y^b,\qquad y=T_m(t).
 \label{eq:ps}
\end{equation}
The blocks are polynomial-division remainders, not consecutive slices of the Chebyshev coefficients. Coefficient transformations are performed offline. Encrypted evaluation forms $t_R(x)$, evaluates blocks using cached Chebyshev values, and combines them using $y,y^2,y^4,\ldots$. The block-zero accumulator includes the exact linear term. No comparison or transcendental evaluation is required.

For the specified dense schedule, the ciphertext--ciphertext multiplication count, including squaring, is
\begin{equation}
 M_{cc}=1+\beta_m+(B-1)+\mu,\quad
 \beta_m=\begin{cases}k-1,&B=1,\\m-1,&B>1,\end{cases}\quad
 \mu=\max\{0,\lceil\log_2B\rceil-1\}.
 \label{eq:cost}
\end{equation}
The terms count the input square, basis and generator products, block combinations, and squares of generator powers. Increasing $m$ increases basis work but reduces the block count. Our splits are fixed rather than optimized for runtime. This schedule-specific count can decrease with pruning. Appendix~\ref{app:algorithms} gives the division algorithm and a complete $k=7,m=4$ example. Plaintext--ciphertext products and consumed modulus levels are recorded separately.

\section{Error bounds for specified coefficients}
\label{sec:analysis}
\subsection{Uniform error and coefficient perturbations}
Theorem~\ref{prop:parity} characterizes the best error over a polynomial space. Deployment requires a bound for the coefficient vector actually evaluated:
\begin{equation}
 \epsilon_A(c,R):=\max_{|x|\le R}|P_c(x)-A(x)|.
 \label{eq:actualerror}
\end{equation}
A finite covering and derivative bound yield the following standard validation argument \citep{chevillard}.

\begin{theorem}[Interval bound and coefficient transfer]
\label{prop:bound}
Let $R>0$, $A\in C^1([-R,R])$, and let $P_c$ be defined by~\eqref{eq:profile}. For ordered points $-R\le x_0<\cdots<x_{n-1}\le R$, $n\ge2$, set
\begin{equation}
 h=\max\left\{x_0+R,\ R-x_{n-1},\
 \frac12\max_{0\le j<n-1}(x_{j+1}-x_j)\right\}.
 \label{eq:cover}
\end{equation}
Suppose $V_j\ge|P_c(x_j)-A(x_j)|$ for all $j$ and $M\ge\|(P_c-A)'\|_{\infty,[-R,R]}$. Then
\begin{equation}
 \epsilon_A(c,R)\le U_c:=\max_jV_j+hM.
 \label{eq:bound}
\end{equation}
Furthermore, if $P_{\widetilde c}$ uses the same radius, Chebyshev basis, and fixed linear term, then
\begin{equation}
 \epsilon_A(\widetilde c,R)\le U_c+\|\widetilde c-c\|_1,
 \label{eq:transfer}
\end{equation}
with coefficient vectors padded by zeros when their lengths differ.
\end{theorem}
Every input is within $h$ of an evaluated point; the mean value theorem gives~\eqref{eq:bound}. The Chebyshev inequality $|T_j(t)|\le1$ gives~\eqref{eq:transfer} (Appendix~\ref{app:validation}). This coefficient bound, unlike the structural reduction, depends on the basis.

Writing $Q_c'(t)=\sum_j c_j^{(1)}T_j(t)$ gives the sufficient derivative bound
\begin{equation}
 M=M_P+M_A,\qquad
 M_P=\frac12+\frac4R\sum_j|c_j^{(1)}|,
 \label{eq:derivative}
\end{equation}
where $M_A\ge\|A'\|_\infty$. Directly enclosing the error derivative may tighten this conservative bound (Appendix~\ref{app:localbound}). A numerical certificate additionally requires outward-rounded enclosures for $V_j$, $M$, and all arithmetic in~\eqref{eq:bound}.

\subsection{Approximation and evaluation error}
Let $A_{\rm fit}$ denote the fitting target and $A_{\rm ref}$ the reference activation. Suppose $U_c$ bounds the approximation to $A_{\rm fit}$, and the stored coefficients are $\widetilde c$. At any $x\in[-R,R]$, the decoded output $Y_{\rm HE}$ obeys
\begin{equation}
 |Y_{\rm HE}-A_{\rm ref}(x)|
 \le \|A_{\rm fit}-A_{\rm ref}\|_{\infty,[-R,R]}
      +U_c+\|\widetilde c-c\|_1
      +|Y_{\rm HE}-P_{\widetilde c}(x)|.
 \label{eq:total}
\end{equation}
The main ciphertext comparisons share fitting and reference targets, so the first term vanishes. The last term includes numerical evaluation and input representation errors. The measured finite-grid discrepancies from a host implementation are not certified bounds over all ciphertexts \citep{ckks}.

Rounding $k+1$ coefficients to $\ell$ decimal places gives $\|\widetilde c-c\|_1\le(k+1)10^{-\ell}/2$ for exact decimal rounding. Binary storage error, when relevant, is additional. Reducing CKKS precision changes a different part of~\eqref{eq:total}. An upper bound exceeding the budget is inconclusive; it does not prove that the actual error exceeds the budget.

\section{Evaluation}
\label{sec:experiments}
\subsection{Experimental setup}
We evaluate frozen-weight Gemma-4-E2B-it and Llama-3.2-3B-Instruct \citep{gemma4,llama32}, separately from ciphertext kernels. Gemma's \emph{Exact-erf} reference replaces tanh-GELU at 35 decoder-MLP sites; \SPA{}, MOAI, and NEXUS approximate erf-GELU at the same sites. \emph{Native-tanh} remains a control. Other gates, attention, normalization, and weights are unchanged. Llama's reference uses native SiLU.

Tasks include multiple-choice questions (MCQ) from HellaSwag, MMLU, Social IQa, and RACE-high, and language modeling on WikiText-2 and C4-en \citep{hellaswag,mmlu,social,race,wikitext,t5}. MCQ choices are ranked by conditional sequence likelihood, not free-form generation. 
Calibration data are disjoint from evaluation data, and calibration is repeated for Exact-erf. The backbone uses BF16; activations are evaluated in FP32 and cast back to BF16. Appendix~\ref{app:protocol} gives splits, range assignment, scoring, and paired statistics. CKKS kernel timings use the Intel Xeon Gold 6530 CPU with a SEAL 4.1-compatible implementation; plaintext model evaluations use the NVIDIA RTX PRO 6000 Blackwell Workstation Edition GPU. Code: \CodeRepositoryLink.

\subsection{Baselines and encrypted operation counts}
\label{sec:baselines}
We compare erf-GELU with MOAI and NEXUS, QuickGELU with Euston, and SiLU with Orion, ENSI, Sylph, and THOR \citep{moai,nexus,euston,orion,ensi,sylph,thor}. MOAI reuses Gemma-erf configurations on $[-5,5]$ (R5) and $[-20,20]$ (R20), including domain extension.

\begin{table*}[ht]
\centering
\caption{Operation counts per 32,768-slot call. CT and PT denote ciphertext and plaintext. CT$\times$CT includes squaring; PT$\times$CT includes normalization and scale alignment. Levels is the input--output chain-index difference. Dashes indicate unevaluated configurations.}
\label{tab:kernelcost}
\small\setlength{\tabcolsep}{5pt}
\begin{tabular}{>{\centering\arraybackslash}m{1.7cm}lrrrrrr}
\toprule
& & \multicolumn{3}{c}{$[-5,5]$} & \multicolumn{3}{c}{$[-20,20]$} \\
\cmidrule(lr){3-5}\cmidrule(lr){6-8}
Activation & Method & CT$\times$CT & PT$\times$CT & Levels & CT$\times$CT & PT$\times$CT & Levels \\
\midrule
 & \SPA{}-fast & 5 & 8 & 6 & 11 & 23 & 7 \\
 & \cellcolor{secblue}\SPA{}-balanced & \cellcolor{secblue}7 & \cellcolor{secblue}11 & \cellcolor{secblue}6 & \cellcolor{secblue}12 & \cellcolor{secblue}30 & \cellcolor{secblue}8 \\
 & \SPA{}-accurate & 7 & 11 & 6 & 14 & 37 & 8 \\
 & MOAI & 23 & 36 & 8 & 29 & 42 & 14 \\
\multirow[c]{-5}{1.7cm}{\centering erf-GELU} & NEXUS & 48 & 41 & 18 & 68 & 57 & 26 \\
\midrule
 & \SPA{}-fast & 5 & 8 & 6 & 11 & 23 & 7 \\
 & \cellcolor{secblue}\SPA{}-balanced & \cellcolor{secblue}7 & \cellcolor{secblue}11 & \cellcolor{secblue}6 & \cellcolor{secblue}14 & \cellcolor{secblue}37 & \cellcolor{secblue}8 \\
 & \SPA{}-accurate & 8 & 14 & 7 & 16 & 51 & 8 \\
\multirow[c]{-4}{1.7cm}{\centering QuickGELU} & Euston & 55 & 43 & 40 & 59 & 43 & 42 \\
\midrule
 & \SPA{}-fast & 3 & 5 & 5 & 9 & 16 & 7 \\
 & \cellcolor{secblue}\SPA{}-balanced & \cellcolor{secblue}5 & \cellcolor{secblue}8 & \cellcolor{secblue}6 & \cellcolor{secblue}11 & \cellcolor{secblue}23 & \cellcolor{secblue}7 \\
 & \SPA{}-accurate & 7 & 11 & 6 & 12 & 30 & 8 \\
 & Orion & 24 & 117 & 9 & 24 & 119 & 9 \\
 & ENSI & 9 & 14 & 7 & -- & -- & -- \\
 & Sylph & 9 & 14 & 7 & -- & -- & -- \\
\multirow[c]{-7}{1.7cm}{\centering SiLU} & THOR & 21 & 38 & 12 & 21 & 38 & 12 \\
\bottomrule
\end{tabular}
\end{table*}

Table~\ref{tab:kernelcost} includes input-mapping costs. 
Methods share a CKKS context within each target and interval, except for QuickGELU: SPA uses a 46-bit scale, whereas Euston uses a separate context with a 35-bit scale. No kernel performs bootstrapping. Appendix~\ref{app:he} details packing and timing.

\subsection{Frozen-model quality}
\begin{table*}[ht]
\centering\small\setlength{\tabcolsep}{4pt}
\caption{Zero-shot strict accuracy (\%); HS/SIQA/RACE denote HellaSwag/Social IQa/RACE-high. Pooled scores are sample-weighted. Changes relative to Exact-erf (Gemma) or Native (Llama) are computed before rounding. Invalid predictions count as incorrect; blue marks balanced.}
\label{tab:mcqmain}
\begin{tabular}{>{\centering\arraybackslash}m{2.8cm}lrrrrrr}
\toprule
Model & Method & HS & MMLU & SIQA & RACE & Pooled & $\Delta$ pp \\
\midrule
 & Native-tanh & 57.76 & 57.61 & 48.46 & 44.00 & 55.44 & -0.04 \\
 & Exact-erf & 57.81 & 57.57 & 48.67 & 44.23 & 55.48 & +0.00 \\
 & \SPA{}-fast & 57.90 & 57.54 & 48.57 & 44.23 & 55.49 & +0.01 \\
 & \cellcolor{secblue}\SPA{}-balanced & \cellcolor{secblue}57.91 & \cellcolor{secblue}57.49 & \cellcolor{secblue}48.52 & \cellcolor{secblue}44.03 & \cellcolor{secblue}55.44 & \cellcolor{secblue}-0.04 \\
 & \SPA{}-accurate & 57.73 & 57.51 & 48.62 & 44.08 & 55.40 & -0.08 \\
 & MOAI & 57.93 & 57.56 & 48.72 & 44.11 & 55.51 & +0.02 \\
\multirow[c]{-7}{2.8cm}{\centering Gemma-4-E2B-it} & NEXUS & 57.42 & 51.69 & 48.16 & 43.85 & 52.47 & -3.01 \\
\midrule
 & Native & 71.65 & 62.16 & 53.33 & 49.60 & 63.31 & +0.00 \\
 & \SPA{}-fast & 71.23 & 61.27 & 51.43 & 49.69 & 62.64 & -0.68 \\
 & \cellcolor{secblue}\SPA{}-balanced & \cellcolor{secblue}71.66 & \cellcolor{secblue}62.36 & \cellcolor{secblue}53.07 & \cellcolor{secblue}49.71 & \cellcolor{secblue}63.41 & \cellcolor{secblue}+0.09 \\
 & \SPA{}-accurate & 71.59 & 62.29 & 53.33 & 49.66 & 63.36 & +0.05 \\
 & Orion & 71.65 & 62.09 & 53.17 & 49.66 & 63.28 & -0.03 \\
 & ENSI & 70.44 & 61.79 & 52.92 & 49.51 & 62.69 & -0.62 \\
 & Sylph & 71.67 & 62.23 & 53.17 & 49.80 & 63.37 & +0.05 \\
\multirow[c]{-8}{2.8cm}{\centering Llama-3.2-3B-Instruct} & THOR & 71.53 & 62.05 & 53.07 & 49.86 & 63.23 & -0.08 \\
\bottomrule
\end{tabular}

\end{table*}
Table~\ref{tab:mcqmain} shows small changes in pooled accuracy for SPA-balanced: $-0.04$ percentage points for Gemma (paired 95\% interval $[-0.1422,+0.0711]$) and $+0.09$ for Llama.

\begin{figure}[ht]
\centering
\begin{minipage}{0.90\linewidth}
\centering
\includegraphics[width=.49\linewidth]{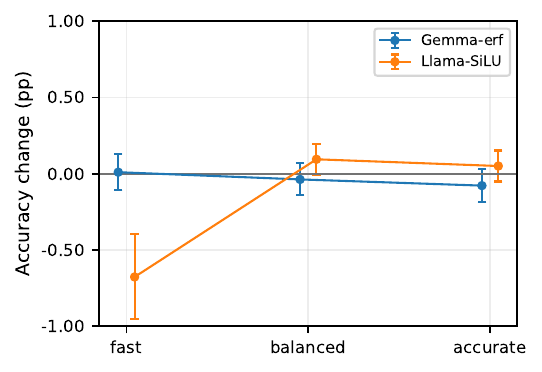}\hfill
\includegraphics[width=.49\linewidth]{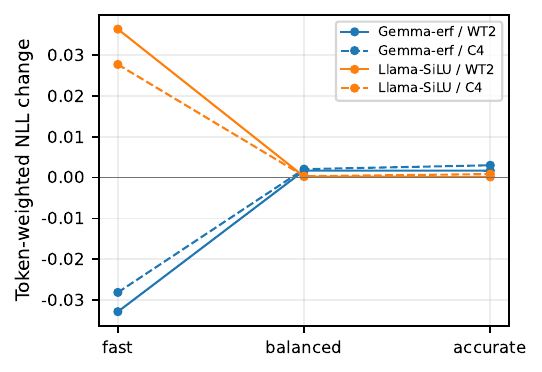}
\end{minipage}
\caption{Model performance changes: pooled MCQ accuracy with paired 95\% confidence intervals (left); token-weighted NLL (right). References: Exact-erf for Gemma, Native for Llama.}
\label{fig:modelresults}
\end{figure}
For Gemma, balanced increases perplexity (PPL) by $0.17\%$ on WikiText-2 and $0.21\%$ on C4-en, versus MOAI's $2.88\%$ and $2.47\%$ (Table~\ref{tab:lmall}). On common windows, $\mathrm{PPL}_m/\mathrm{PPL}_{\rm ref}=e^{\Delta\mathrm{NLL}}$ links these values to Figure~\ref{fig:modelresults}. NEXUS yields 1,609 invalid MCQ predictions and fails on 103/368 WikiText-2 windows and 534/1,501 C4-en windows. Gains on shared windows therefore do not imply full-corpus gains.

\subsection{Ciphertext accuracy and latency}
\begin{figure}[ht]
    \centering
    \includegraphics[width=0.90\linewidth]{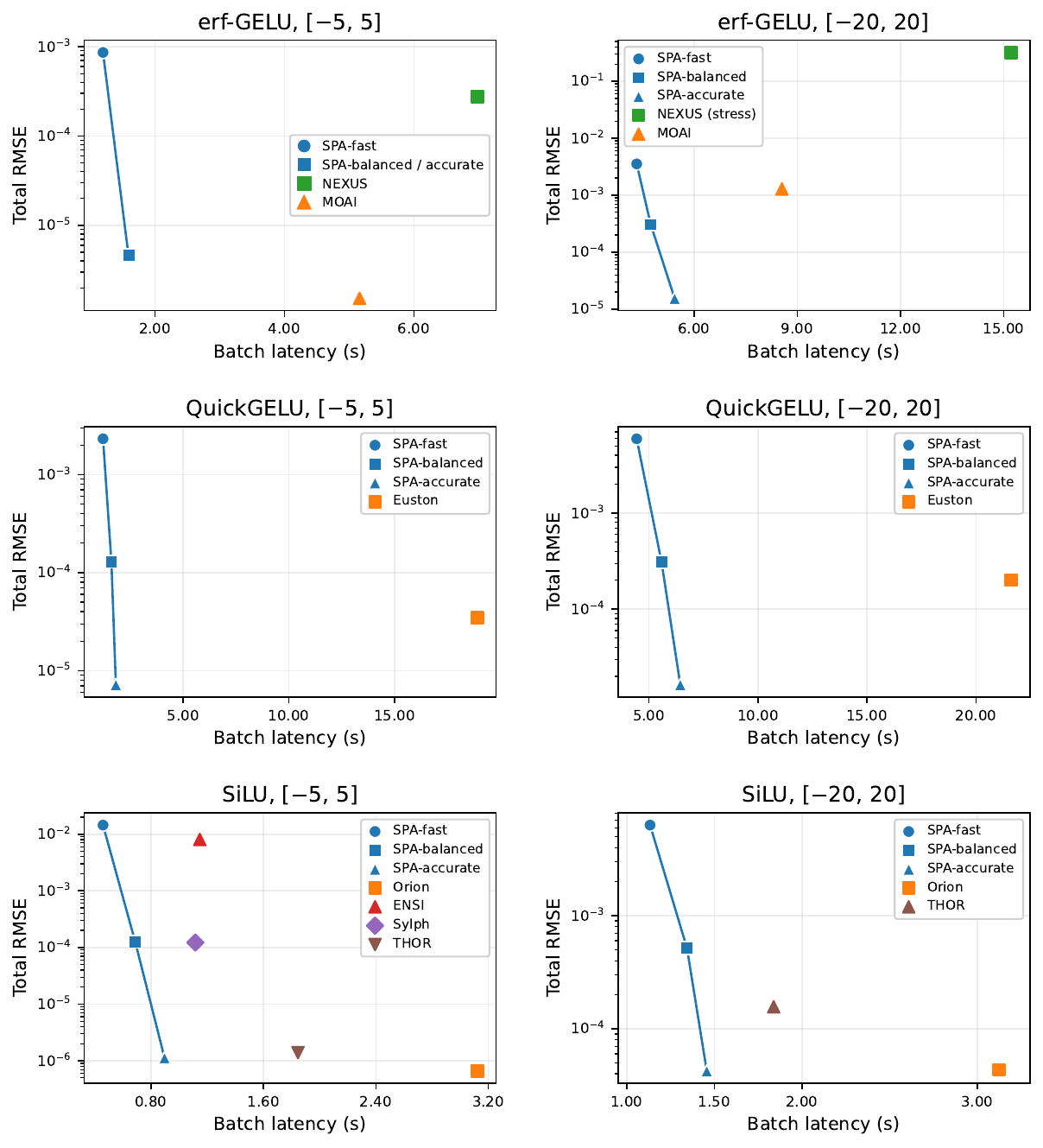}
    \caption{Ciphertext error--latency trade-offs. Rows show erf-GELU, QuickGELU, and SiLU; columns show $[-5,5]$ and $[-20,20]$. Blue lines connect \SPA{} variants. Each panel uses a common target.}
    \label{fig:kernel_all}
\label{fig:kernels}
\end{figure}
\begin{table*}[ht]
\centering
\caption{Ciphertext results on a common 32,768-point grid: median batch latency and output errors. Scale: 35 bits for Euston, 46 for other methods. $^\dagger$NEXUS R20 is a range-extension stress test.}
\label{tab:hemain}
\small\setlength{\tabcolsep}{6pt}
\begin{tabular}{>{\centering\arraybackslash}m{1.8cm}llrcc}
\toprule
Activation & Interval & Method & Time (s) & RMSE & Max. abs. error \\
\midrule
 & $[-5,5]$ & \SPA{}-fast & 1.20 & $8.61\!\times\!10^{-4}$ & $1.22\!\times\!10^{-3}$ \\
 & \cellcolor{secblue}$[-5,5]$ & \cellcolor{secblue}\SPA{}-balanced & \cellcolor{secblue}1.59 & \cellcolor{secblue}$4.63\!\times\!10^{-6}$ & \cellcolor{secblue}$7.19\!\times\!10^{-6}$ \\
 & $[-5,5]$ & \SPA{}-accurate & 1.60 & $4.63\!\times\!10^{-6}$ & $7.19\!\times\!10^{-6}$ \\
 & $[-5,5]$ & MOAI & 5.16 & $1.53\!\times\!10^{-6}$ & $3.81\!\times\!10^{-6}$ \\
 & $[-5,5]$ & NEXUS & 6.98 & $2.76\!\times\!10^{-4}$ & $6.90\!\times\!10^{-4}$ \\
 & $[-20,20]$ & \SPA{}-fast & 4.33 & $3.54\!\times\!10^{-3}$ & $5.01\!\times\!10^{-3}$ \\
 & \cellcolor{secblue}$[-20,20]$ & \cellcolor{secblue}\SPA{}-balanced & \cellcolor{secblue}4.74 & \cellcolor{secblue}$3.06\!\times\!10^{-4}$ & \cellcolor{secblue}$4.37\!\times\!10^{-4}$ \\
 & $[-20,20]$ & \SPA{}-accurate & 5.43 & $1.55\!\times\!10^{-5}$ & $2.56\!\times\!10^{-5}$ \\
 & $[-20,20]$ & MOAI & 8.55 & $1.29\!\times\!10^{-3}$ & $4.54\!\times\!10^{-3}$ \\
\multirow[c]{-10}{1.8cm}{\centering erf-GELU} & $[-20,20]$ & NEXUS$^\dagger$ & 15.21 & $3.20\!\times\!10^{-1}$ & $6.49\!\times\!10^{0}$ \\
\midrule
 & $[-5,5]$ & \SPA{}-fast & 1.21 & $2.32\!\times\!10^{-3}$ & $3.29\!\times\!10^{-3}$ \\
 & \cellcolor{secblue}$[-5,5]$ & \cellcolor{secblue}\SPA{}-balanced & \cellcolor{secblue}1.61 & \cellcolor{secblue}$1.29\!\times\!10^{-4}$ & \cellcolor{secblue}$1.83\!\times\!10^{-4}$ \\
 & $[-5,5]$ & \SPA{}-accurate & 1.82 & $7.19\!\times\!10^{-6}$ & $1.08\!\times\!10^{-5}$ \\
 & $[-5,5]$ & Euston & 18.91 & $3.46\!\times\!10^{-5}$ & $4.67\!\times\!10^{-4}$ \\
 & $[-20,20]$ & \SPA{}-fast & 4.44 & $5.94\!\times\!10^{-3}$ & $8.41\!\times\!10^{-3}$ \\
 & \cellcolor{secblue}$[-20,20]$ & \cellcolor{secblue}\SPA{}-balanced & \cellcolor{secblue}5.58 & \cellcolor{secblue}$3.11\!\times\!10^{-4}$ & \cellcolor{secblue}$4.44\!\times\!10^{-4}$ \\
 & $[-20,20]$ & \SPA{}-accurate & 6.44 & $1.64\!\times\!10^{-5}$ & $2.74\!\times\!10^{-5}$ \\
\multirow[c]{-8}{1.8cm}{\centering QuickGELU} & $[-20,20]$ & Euston & 21.61 & $2.01\!\times\!10^{-4}$ & $2.92\!\times\!10^{-3}$ \\
\midrule
 & $[-5,5]$ & \SPA{}-fast & 0.46 & $1.44\!\times\!10^{-2}$ & $2.05\!\times\!10^{-2}$ \\
 & \cellcolor{secblue}$[-5,5]$ & \cellcolor{secblue}\SPA{}-balanced & \cellcolor{secblue}0.69 & \cellcolor{secblue}$1.26\!\times\!10^{-4}$ & \cellcolor{secblue}$1.79\!\times\!10^{-4}$ \\
 & $[-5,5]$ & \SPA{}-accurate & 0.90 & $1.13\!\times\!10^{-6}$ & $2.24\!\times\!10^{-6}$ \\
 & $[-5,5]$ & Orion & 3.12 & $6.61\!\times\!10^{-7}$ & $1.22\!\times\!10^{-6}$ \\
 & $[-5,5]$ & ENSI & 1.15 & $8.03\!\times\!10^{-3}$ & $2.09\!\times\!10^{-2}$ \\
 & $[-5,5]$ & Sylph & 1.12 & $1.23\!\times\!10^{-4}$ & $1.72\!\times\!10^{-4}$ \\
 & $[-5,5]$ & THOR & 1.85 & $1.40\!\times\!10^{-6}$ & $3.61\!\times\!10^{-6}$ \\
 & $[-20,20]$ & \SPA{}-fast & 1.13 & $6.32\!\times\!10^{-3}$ & $8.94\!\times\!10^{-3}$ \\
 & \cellcolor{secblue}$[-20,20]$ & \cellcolor{secblue}\SPA{}-balanced & \cellcolor{secblue}1.34 & \cellcolor{secblue}$5.17\!\times\!10^{-4}$ & \cellcolor{secblue}$7.35\!\times\!10^{-4}$ \\
 & $[-20,20]$ & \SPA{}-accurate & 1.46 & $4.23\!\times\!10^{-5}$ & $6.37\!\times\!10^{-5}$ \\
 & $[-20,20]$ & Orion & 3.12 & $4.33\!\times\!10^{-5}$ & $1.67\!\times\!10^{-4}$ \\
\multirow[c]{-12}{1.8cm}{\centering SiLU} & $[-20,20]$ & THOR & 1.84 & $1.56\!\times\!10^{-4}$ & $3.40\!\times\!10^{-4}$ \\
\bottomrule
\end{tabular}
\end{table*}

For erf-GELU, balanced achieves a $3.25\times$ speedup over MOAI on R5 (paired 95\% confidence interval: $[3.2169,3.2572]$; Table~\ref{tab:hemain}). Both methods have RMSEs on the $10^{-6}$ scale, although balanced has $3.04$ times MOAI's RMSE. On R20, its RMSE is $76.23\%$ lower with a $1.80\times$ speedup. MOAI’s domain extension adds cost. On R5, SPA-balanced and SPA-accurate share the same polynomial and evaluation graph.

For QuickGELU, accurate achieves $79.22\%$ lower RMSE and a $10.40\times$ speedup over Euston on $[-5,5]$; on R20, RMSE is $91.84\%$ lower with a $3.36\times$ speedup. Balanced is faster but has higher RMSE and lower maximum error than Euston. Their CKKS contexts differ. For SiLU R5, accurate achieves $19.02\%$ lower RMSE with a $2.06\times$ speedup over THOR. Balanced's $2.68\times$ speedup over THOR trades accuracy for latency; against Sylph, its RMSE is $2.48\%$ higher with a $1.62\times$ speedup. R20 includes only Orion and THOR.

\subsection{Coefficient and CKKS precision}
\begin{figure}[ht]
\centering
\begin{minipage}{0.90\linewidth}
\centering
\includegraphics[width=.49\linewidth]{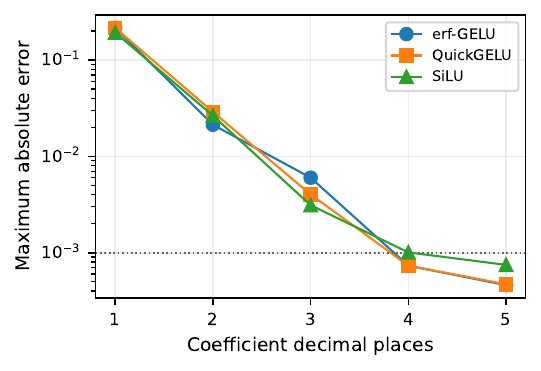}\hfill
\includegraphics[width=.49\linewidth]{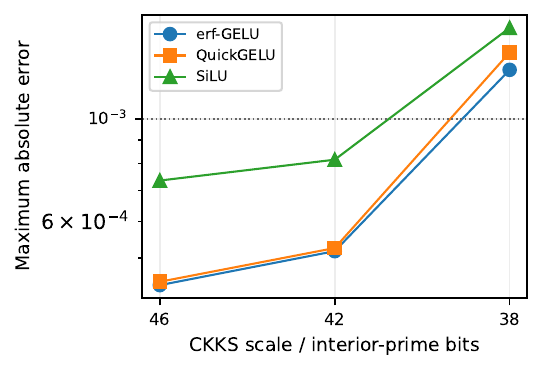}
\end{minipage}
\caption{Precision effects for balanced R20 configurations: residual-coefficient rounding (left) and CKKS scale and intermediate-prime bit sizes (right).}
\label{fig:precision}
\end{figure}

Rounding coefficients to three decimal places gives a maximum absolute error above $10^{-3}$ on the test grid. Four decimals meet this budget for erf-GELU and QuickGELU; SiLU reaches $1.01\times10^{-3}$. Unrounded coefficients meet it at 46/42-bit CKKS precision, but not 38 bits. These are sampled observations, not uniform guarantees.

\FloatBarrier
\section{Limitations and conclusion}
\label{sec:discussion}
\SPA{} preserves the known odd part and best uniform approximation error for activations with the structure in~\eqref{eq:class}. The measured error--cost trade-offs do not establish superiority over all optimized even-polynomial evaluators. Model results are empirical, with task-specific calibration in two LLMs; Gemma uses Exact-erf. Scalar guarantees exclude unverified configurations and out-of-range inputs. Kernel speedups do not establish end-to-end gains.

\FloatBarrier
\label{lastmainpage}
\clearpage
\section*{Reproducibility Statement}
Sections~\ref{sec:method}--\ref{sec:analysis} define the approximation and its assumptions. Appendix~\ref{app:proofs} provides proofs, Appendix~\ref{app:validation} records numerical verification status, and Appendix~\ref{app:algorithms} specifies coefficient generation and polynomial evaluation. Data processing, calibration, and paired evaluation are described in Appendix~\ref{app:protocol}. The accompanying material includes the stored coefficients, original result records, and scripts for regenerating the numerical tables and figures. Code repository: \CodeRepositoryLink.

\paragraph{AI use statement.}
Generative AI tools assisted with code organization and standardization and with refining the manuscript's language and wording.

% Use the exact style and bibliography files supplied by the author.
\bibliographystyle{iclr2027_conference}
\bibliography{references}
\clearpage
\appendix
\section{Proofs and activation-specific interpretation}
\label{app:proofs}
\subsection{Proof of Theorem~\ref{prop:parity}}
Let $g$ and $A$ satisfy the assumptions of the theorem. The structure in~\eqref{eq:class} gives
\[
 A(-x)=(-x)g(-x)=-x[1-g(x)]=A(x)-x.
\]
Consequently, $r_A(x)=A(x)-x/2$ is even, and the continuous function $H_A(u)=r_A(\sqrt u)$ satisfies $H_A(x^2)=r_A(x)$ on $[-R,R]$. This identity explains both the definition of $H_A$ and its domain $[0,R^2]$.

For an arbitrary $p\in\Pi_{2k+1}$, define the structure-preserving projection
\[
 (\mathcal Sp)(x)=\frac{x}{2}+\frac{p(x)+p(-x)}{2}.
\]
Writing $e=A-p$ and using $A(-x)=A(x)-x$ yields
\[
 A(x)-(\mathcal Sp)(x)=\frac{e(x)+e(-x)}{2},
 \qquad
 \|A-\mathcal Sp\|_{\infty,[-R,R]}\le\|A-p\|_{\infty,[-R,R]}.
\]
The even part of $p$ has degree at most $2k$, so it equals $q(x^2)$ for some $q\in\Pi_k$. For $k\ge1$, the polynomial $\mathcal Sp=x/2+q(x^2)$ belongs to $\Pi_{2k}$. Thus every candidate in the larger space admits a structure-preserving replacement with no greater uniform error. Conversely, every structured candidate belongs to $\Pi_{2k}\subset\Pi_{2k+1}$. Finally, $x\mapsto x^2$ maps $[-R,R]$ onto $[0,R^2]$, so
\[
 \left\|A-\left(\frac{x}{2}+q(x^2)\right)\right\|_{\infty,[-R,R]}
 =\|H_A-q\|_{\infty,[0,R^2]}.
\]
Taking infima proves~\eqref{eq:parity}. Reflection and differentiation of $P_c=x/2+Q_c(t_R(x))$ give~\eqref{eq:invariants}, since $t_R$ is even and $t_R'(0)=0$. Changing $c$ preserves both identities in real arithmetic; separately rounded floating-point or ciphertext outputs need not satisfy them exactly.\hfill$\square$

\subsection{Regularity at the origin}
\label{app:analytic}
Theorem~\ref{prop:parity} assumes only continuity. Under the additional assumption that $g$ is analytic near zero, the identity $g(-x)=1-g(x)$ gives a convergent series
\[
 g(x)=\frac12+\sum_{j\ge0}a_{2j+1}x^{2j+1},
 \qquad H_A(u)=\sum_{j\ge0}a_{2j+1}u^{j+1}.
\]
Thus the square-root expression for $H_A$ has no singularity at zero for these analytic gates. This applies locally to the sigmoid, erf, and tanh-based gates used here; it does not assert that all residuals are entire. Continuity and parity alone are insufficient: on $[-1,1]$, take $g(x)=1/2+\operatorname{sgn}(x)|x|^{1/2}/4$, with $g(0)=1/2$. Then $H_A(u)=u^{3/4}/4$, which is continuous but not differentiable at zero.

\subsection{Residuals of the activation examples}
\label{app:spectra}
The following identities characterize the residuals; they are not the procedures used to generate the reported numerical coefficients. The SiLU residual is
\[
 H_{\rm S}(u)=\frac{\sqrt u}{2}\tanh(\sqrt u/2)
 =\sum_{n=0}^{\infty}\frac{2u}{u+\lambda_n},
 \qquad\lambda_n=\pi^2(2n+1)^2,
\]
by the classical hyperbolic partial-fraction expansion \citep{silu,dlmf}. With $S=R^2$, the classical minimax formula for a reciprocal \citep{kvz} gives the best uniform error for a degree-at-most-$k$ approximation to the $n$th summand:
\[
 e_{n,k}=\frac{S}{S+\lambda_n}\rho_n^k,
 \qquad
 \rho_n=\frac{\sqrt{S+\lambda_n}-\sqrt{\lambda_n}}
 {\sqrt{S+\lambda_n}+\sqrt{\lambda_n}}.
\]
If $\tau_k=\sum_{n\ge1}e_{n,k}$, the distance to the polynomial subspace satisfies
\[
 \max\{0,e_{0,k}-\tau_k\}
 \le \inf_{q\in\Pi_k}\|H_{\rm S}-q\|_{\infty,[0,S]}
 \le e_{0,k}+\tau_k.
\]
The upper bound follows by summing the individual approximants: the function series converges uniformly, the error bounds are summable, and the approximant sum lies in the fixed finite-dimensional polynomial space. The lower bound follows from the reverse triangle inequality for distance to that subspace. These bounds concern best-approximation errors and the indicated analytic construction; they do not directly bound arbitrary stored coefficient vectors.

QuickGELU with $\beta=1.702$ satisfies the direct scaling identities
\[
 A_{\rm Q}(x)=\beta^{-1}A_{\rm S}(\beta x),\qquad
 H_{\rm Q}(u)=\beta^{-1}H_{\rm S}(\beta^2u)
 =\frac{\sqrt u}{2}\tanh(\beta\sqrt u/2).
\]
The experiments nevertheless fit QuickGELU directly on each stated interval; SiLU coefficients from a different interval are not reused without the required scaling.

For erf-GELU,
\[
 H_{\rm E}(u)=\frac{u}{\sqrt{2\pi}}\int_0^1 e^{-uv^2/2}\,dv.
\]
The right-hand side extends to an entire function of $u$. On a fixed interval, analyticity on every Bernstein ellipse implies polynomial approximation errors that asymptotically decay faster than any fixed geometric rate, but does not order the finite-degree errors for different targets \citep{trefethen}. This is the target used by the Exact-erf Gemma reference and its polynomial replacements, rather than the original Gemma activation.

For the original tanh-GELU, write $a=\sqrt{2/\pi}$ and $b=0.044715$. Its residual has the representation
\[
 H_{\rm T}(u)=\sum_{n\ge0}
 \frac{u(1+bu)}{a\{u(1+bu)^2+d_n\}},
 \qquad d_n=\left[\frac{\pi(n+1/2)}{a}\right]^2.
\]
The cubic preimage equation for these poles also appears in \citet{tang2026}. With $y=bu$ and $\delta=bd_n$, the denominator equation becomes $y(1+y)^2+\delta=0$, whose discriminant is $\delta(4-27\delta)$. The resulting pole locations differ from the negative real poles in the SiLU representation. One cannot substitute the SiLU reciprocal error formula into tanh-GELU without accounting for those differences. This explanatory observation is separate from the numerical fitting algorithm.

\section{Details of coefficient-specific error validation}
\label{app:validation}
\subsection{Proof of Theorem~\ref{prop:bound}}
Let $e=P_c-A$. For every $x\in[-R,R]$, one of the evaluated points $x_j$ is at distance at most $h$ as defined in~\eqref{eq:cover}: the endpoint distances cover the boundary segments, and half the distance between adjacent points bounds the distance to the nearer point. The mean value theorem gives
\[
 |e(x)|\le|e(x_j)|+\|e'\|_\infty|x-x_j|
 \le\max_j V_j+hM.
\]
Taking the maximum proves~\eqref{eq:bound}. For a second coefficient vector, pad both by zeros to a common length. The bound $|T_j(t)|\le1$ gives
\[
 |P_{\widetilde c}(x)-P_c(x)|
 \le\sum_j|\widetilde c_j-c_j|\,|T_j(t_R(x))|
 \le\|\widetilde c-c\|_1.
\]
Combining this with~\eqref{eq:bound} proves~\eqref{eq:transfer}.\hfill$\square$

If $Q_c'(t)=\sum_jc_j^{(1)}T_j(t)$, then $|t_R'(x)|\le4/R$ gives~\eqref{eq:derivative}. For SiLU, with $s=|x|$,
\[
 A'_{\rm S}(x)=\sigma(x)+\frac{x}{2+2\cosh x},\qquad
 \frac{s}{2+2\cosh s}\le\frac{s}{4+s^2}\le\frac14.
\]
Thus $\|A'_{\rm S}\|_\infty\le1.25$, a sufficient bound rather than the sharp maximum. The QuickGELU scaling above implies $A'_{\rm Q}(x)=A'_{\rm S}(\beta x)$, giving the same bound. For erf-GELU a finite derivative bound exists, but the current validation implementation does not provide an interval enclosure for the erf evaluations; the stored result is therefore a sampled check, not a certificate.

\subsection{Possible refinement of the derivative bound}
\label{app:localbound}
Theorem~\ref{prop:bound} accepts any valid enclosure of $e'=P_c'-A'$. The sum $M_P+M_A$ is a convenient sufficient choice, not a requirement. A local alternative partitions $[-R,R]$ into intervals $I_j$ with midpoint $z_j$ and half-width $h_j$. If $V_j\ge|e(z_j)|$ and $D_j\ge\sup_{x\in I_j}|e'(x)|$, then
\[
 \|e\|_{\infty,[-R,R]}\le\max_j(V_j+h_jD_j).
\]
This follows by the mean value theorem within each interval. Direct interval evaluation of $P_c'-A'$ may exploit cancellation and reduce overestimation, but a sampled derivative maximum is not a valid replacement for $D_j$. This refinement has not been used to generate Table~\ref{tab:validationstatus}. The missing erf-GELU certificate is caused by the absence of an erf enclosure in the current validator, not by a demonstrated failure of the derivative estimate.

\subsection{Arithmetic assumptions and current status}
The validation uses the actual float64 evaluation points and accounts for endpoint distances and gaps between processing chunks when computing $h$. Its interval arithmetic uses NumPy long double with \texttt{nmant=63}, assuming correctly rounded basic operations in the relevant normal range and outward-enclosed constants. The exponential enclosure reduces $z$ as $z=n\ln2+r$ and checks $|r|<0.69<\ln2$. A degree-24 Taylor expansion has the explicit remainder
\[
 \left|e^r-\sum_{j=0}^{24}\frac{r^j}{j!}\right|
 \le\frac{2|r|^{25}}{25!}.
\]
The implementation expands computed endpoints outward with \texttt{nextafter}, under its stated correct-rounding and range assumptions; it does not rely on round-to-nearest arithmetic alone. Outward evaluation and final scalar conversion are part of those assumptions. Porting requires rechecking them. An independent ball-arithmetic validation, for example with Arb \citep{arb}, could reduce platform dependence and supply an erf enclosure, but has not been performed here.

\begin{table}[ht]
\centering\small\setlength{\tabcolsep}{4pt}
\caption{Validation status of the balanced ciphertext coefficients. Interval bounds are rounded upward for display. A dash indicates that no continuous-interval bound was produced, not that the error is zero.}
\label{tab:validationstatus}
\begin{tabular}{lrrrll}
\toprule
Target & $R$ & $k$ & Grid max. error & Interval upper bound & Status \\
\midrule
SiLU & 5 & 7 & $1.78\!\times\!10^{-4}$ & $1.96\!\times\!10^{-4}$ & Interval bound \\
SiLU & 20 & 23 & $7.31\!\times\!10^{-4}$ & $8.05\!\times\!10^{-4}$ & Interval bound \\
erf-GELU & 5 & 11 & $6.60\!\times\!10^{-6}$ & -- & Sampled only \\
erf-GELU & 20 & 31 & $4.33\!\times\!10^{-4}$ & -- & Sampled only \\
QuickGELU & 5 & 11 & $1.83\!\times\!10^{-4}$ & $2.02\!\times\!10^{-4}$ & Interval bound \\
QuickGELU & 20 & 39 & $4.40\!\times\!10^{-4}$ & $4.85\!\times\!10^{-4}$ & Interval bound \\
\bottomrule
\end{tabular}

\end{table}
Table~\ref{tab:validationstatus} distinguishes validated SiLU/QuickGELU bounds from the sampled erf-GELU checks. The available results do not include a completed coefficient-transfer certificate for the Gemma-erf vectors.

\section{Plaintext evaluation protocol}
\label{app:protocol}

Calibration and range assignment for Gemma are performed on the Exact-erf model, not copied from the original tanh trajectory. Each method independently replaces the activation at the same sites; its approximation is applied to the original activation input, not to an erf-GELU output. The static plan and coefficients are frozen before evaluation. Test labels are not used to choose them.

\begin{table}[ht]
\centering\small\setlength{\tabcolsep}{4pt}
\caption{Calibration and evaluation splits. Calibration counts are examples or text windows and include all candidate MCQ continuations; language-modeling counts are windows, not independent documents.}
\begin{tabular}{lrlrll}
\toprule
Task & Calib. $n$ & Split & Test $n$ & Split & Score \\
\midrule
HellaSwag & 39905 & train & 10042 & validation & bytes \\
MMLU & 1816 & validation,dev & 14042 & test & raw \\
Social IQa & 33410 & train & 1954 & validation & bytes \\
RACE-high & 8000 & train & 3498 & test & bytes \\
WikiText-2 & 1500 & train & 368 & test & token NLL \\
C4-en & 1500 & train & 1501 & test & token NLL \\
\bottomrule
\end{tabular}

\end{table}

\paragraph{Inputs and scoring.}
The main results use zero-shot conditional sequence scores with plain prompts. HellaSwag and Social IQa have 512-token input budgets, MMLU has 1,024, and RACE-high has 2,048. MMLU uses raw continuation log-likelihood; the other three tasks divide by the UTF-8 byte length of the continuation, including the separator. Padding is excluded. This scoring protocol is not intended to reproduce public leaderboard results obtained with chat templates.

\paragraph{Paired statistics and failures.}
For strict correctness $a_{m,i}\in\{0,1\}$, the accuracy difference is $100N^{-1}\sum_i(a_{m,i}-a_{0,i})$. Invalid predictions count as incorrect. Pooled accuracy weights individual examples; macro accuracy weights tasks equally. Task-specific random streams derived using SHA256 are independent across tasks; methods within each task share resampling indices. Intervals use 10,000 paired percentile-bootstrap replicates conditional on the evaluated examples.

For shared valid language-modeling windows $J$, with total negative log-likelihood (NLL) $L_{m,i}$ and scored-token count $n_i$, we use
\[
 \Delta_{\rm NLL}=\frac{\sum_{i\in J}(L_{m,i}-L_{0,i})}{\sum_{i\in J}n_i},
 \qquad
 \Delta_{\rm PPL}(\%)=100\{e^{\Delta_{\rm NLL}}-1\}.
\]

\begin{table}[ht]
\centering\small\setlength{\tabcolsep}{4pt}
\caption{Static range assignments. Percentages use all channel-within-layer assignments as denominator, including the separate wide group. Wide channels are not layers or tokens; their small share rounds to zero at two decimals. These are not runtime coverage rates.}
\label{tab:ranges}
\begin{tabular}{llrrrr}
\toprule
Model/target & Task & R5 (\%) & R20 (\%) & $R_{\rm wide}$ & Wide channels \\
\midrule
Gemma-erf & HellaSwag & 86.54 & 13.46 & 32 & 3 \\
Gemma-erf & MMLU & 89.90 & 10.10 & 32 & 3 \\
Gemma-erf & Social IQa & 90.32 & 9.68 & 32 & 2 \\
Gemma-erf & RACE-high & 86.58 & 13.42 & 32 & 5 \\
Gemma-erf & WikiText-2 & 87.62 & 12.38 & 32 & 3 \\
Gemma-erf & C4-en & 86.90 & 13.10 & 32 & 3 \\
Llama-SiLU & HellaSwag & 93.84 & 6.16 & 40 & 3 \\
Llama-SiLU & MMLU & 96.49 & 3.51 & 48 & 3 \\
Llama-SiLU & Social IQa & 96.33 & 3.67 & 48 & 3 \\
Llama-SiLU & RACE-high & 93.58 & 6.42 & 56 & 3 \\
Llama-SiLU & WikiText-2 & 95.03 & 4.97 & 80 & 3 \\
Llama-SiLU & C4-en & 93.95 & 6.05 & 80 & 3 \\
\bottomrule
\end{tabular}

\end{table}
Each assignment is one scalar channel at a particular decoder-MLP activation location, reused across token positions. The R5, R20, and wide groups are disjoint. If their counts are $N_5,N_{20},N_w$, Table~\ref{tab:ranges} reports $100N_5/N$ and $100N_{20}/N$, where $N=N_5+N_{20}+N_w$, and reports $N_w$ as an exact count. For example, Gemma-erf/HellaSwag has $292{,}421$, $45{,}496$, and $3$ assignments, respectively, out of $337{,}920$. Rounding explains why the first two displayed percentages sum to $100\%$.

\section{Ciphertext settings and complete kernel results}
\label{app:he}
The implementation uses a SEAL-compatible CKKS 4.1 interface, ring degree $N=65{,}536$, and $N/2=32{,}768$ occupied slots. The timings in Table~\ref{tab:hemain} and Figure~\ref{fig:kernels} are CPU measurements, not GPU or end-to-end model latencies.

\begin{table}[ht]
\centering\small\setlength{\tabcolsep}{5pt}
\caption{Context parameters for the main runs. All contexts use $N=65,536$. Total modulus bits include the special prime. Euston uses a separate chain; other methods share a context within each target and interval.}
\begin{tabular}{lr lrrr}
\toprule
Target & $R$ & Context & Scale bits & Total modulus bits & Input index \\
\midrule
erf-GELU & 5 & Common & 46 & 1132 & 22 \\
erf-GELU & 20 & Common & 46 & 1500 & 30 \\
QuickGELU & 5 & Common & 46 & 1132 & 22 \\
QuickGELU & 5 & Euston & 35 & 1660 & 44 \\
QuickGELU & 20 & Common & 46 & 1500 & 30 \\
QuickGELU & 20 & Euston & 35 & 1730 & 46 \\
SiLU & 5 & Common & 46 & 856 & 16 \\
SiLU & 20 & Common & 46 & 856 & 16 \\
\bottomrule
\end{tabular}

\end{table}

\begin{table}[ht]
\centering\small\setlength{\tabcolsep}{4pt}
\caption{\SPA{} polynomial degrees and evaluation splits used in the experiments. $k$ is the degree bound in the polynomial variable $t$; $2k$ is the corresponding upper bound in the original input $x$.}
\label{tab:degrees}
\begin{tabular}{lrlrrrr}
\toprule
Target & $R$ & Variant & $\deg Q\le k$ & $\deg P\le2k$ & $m$ & Blocks \\
\midrule
erf-GELU & 5 & \SPA{}-fast & 7 & 14 & 4 & 2 \\
erf-GELU & 5 & \SPA{}-balanced & 11 & 22 & 4 & 3 \\
erf-GELU & 5 & \SPA{}-accurate & 11 & 22 & 4 & 3 \\
erf-GELU & 20 & \SPA{}-fast & 23 & 46 & 8 & 3 \\
erf-GELU & 20 & \SPA{}-balanced & 31 & 62 & 8 & 4 \\
erf-GELU & 20 & \SPA{}-accurate & 39 & 78 & 8 & 5 \\
QuickGELU & 5 & \SPA{}-fast & 7 & 14 & 4 & 2 \\
QuickGELU & 5 & \SPA{}-balanced & 11 & 22 & 4 & 3 \\
QuickGELU & 5 & \SPA{}-accurate & 15 & 30 & 4 & 4 \\
QuickGELU & 20 & \SPA{}-fast & 23 & 46 & 8 & 3 \\
QuickGELU & 20 & \SPA{}-balanced & 39 & 78 & 8 & 5 \\
QuickGELU & 20 & \SPA{}-accurate & 55 & 110 & 8 & 7 \\
SiLU & 5 & \SPA{}-fast & 3 & 6 & 4 & 1 \\
SiLU & 5 & \SPA{}-balanced & 7 & 14 & 4 & 2 \\
SiLU & 5 & \SPA{}-accurate & 11 & 22 & 4 & 3 \\
SiLU & 20 & \SPA{}-fast & 15 & 30 & 8 & 2 \\
SiLU & 20 & \SPA{}-balanced & 23 & 46 & 8 & 3 \\
SiLU & 20 & \SPA{}-accurate & 31 & 62 & 8 & 4 \\
\bottomrule
\end{tabular}

\end{table}
For the recorded dense schedule, with $B=\lceil(k+1)/m\rceil$, the depth model is
\[
 D_{\rm rs}=2+\eta(k,m)+\lceil\log_2(B+1)\rceil,
 \quad
 \eta(k,m)=\begin{cases}\lceil\log_2k\rceil,&B=1,\\\lceil\log_2m\rceil,&B>1.\end{cases}
\]

\small\begin{longtable}{lrrrrrr}\caption{All main ciphertext configurations. Errors $E$ are grid maxima; RMSE refers to decoded output versus the common target. $E_{\rm target}=E_{\rm coeff}=0$ for these unrounded, target-matched runs. $E_{\rm HE}$ is the discrepancy from the host implementation, not a certified CKKS noise bound.}\label{tab:heall}\\
\toprule Method & Time & $E_{\rm poly}$ & $E_{\rm HE}$ & $E_{\rm total}$ & RMSE & Levels \\ \midrule\endfirsthead
\toprule Method & Time & $E_{\rm poly}$ & $E_{\rm HE}$ & $E_{\rm total}$ & RMSE & Levels \\ \midrule\endhead
\multicolumn{7}{l}{\textit{erf-GELU, $R=5$}} \\
\SPA{}-fast & 1.20 & $1.22\!\times\!10^{-3}$ & $6.89\!\times\!10^{-7}$ & $1.22\!\times\!10^{-3}$ & $8.61\!\times\!10^{-4}$ & 6 \\
\SPA{}-balanced & 1.59 & $6.60\!\times\!10^{-6}$ & $6.69\!\times\!10^{-7}$ & $7.19\!\times\!10^{-6}$ & $4.63\!\times\!10^{-6}$ & 6 \\
\SPA{}-accurate & 1.60 & $6.60\!\times\!10^{-6}$ & $6.69\!\times\!10^{-7}$ & $7.19\!\times\!10^{-6}$ & $4.63\!\times\!10^{-6}$ & 6 \\
MOAI & 5.16 & $3.81\!\times\!10^{-6}$ & $1.21\!\times\!10^{-8}$ & $3.81\!\times\!10^{-6}$ & $1.53\!\times\!10^{-6}$ & 8 \\
NEXUS & 6.98 & $6.88\!\times\!10^{-4}$ & $3.30\!\times\!10^{-6}$ & $6.90\!\times\!10^{-4}$ & $2.76\!\times\!10^{-4}$ & 18 \\
\midrule
\multicolumn{7}{l}{\textit{erf-GELU, $R=20$}} \\
\SPA{}-fast & 4.33 & $5.01\!\times\!10^{-3}$ & $4.64\!\times\!10^{-6}$ & $5.01\!\times\!10^{-3}$ & $3.54\!\times\!10^{-3}$ & 7 \\
\SPA{}-balanced & 4.74 & $4.33\!\times\!10^{-4}$ & $4.49\!\times\!10^{-6}$ & $4.37\!\times\!10^{-4}$ & $3.06\!\times\!10^{-4}$ & 8 \\
\SPA{}-accurate & 5.43 & $2.18\!\times\!10^{-5}$ & $4.52\!\times\!10^{-6}$ & $2.56\!\times\!10^{-5}$ & $1.55\!\times\!10^{-5}$ & 8 \\
MOAI & 8.55 & $4.54\!\times\!10^{-3}$ & $4.69\!\times\!10^{-8}$ & $4.54\!\times\!10^{-3}$ & $1.29\!\times\!10^{-3}$ & 14 \\
NEXUS & 15.21 & $6.96\!\times\!10^{-4}$ & $6.49\!\times\!10^{0}$ & $6.49\!\times\!10^{0}$ & $3.20\!\times\!10^{-1}$ & 26 \\
\midrule
\multicolumn{7}{l}{\textit{QuickGELU, $R=5$}} \\
\SPA{}-fast & 1.21 & $3.29\!\times\!10^{-3}$ & $7.22\!\times\!10^{-7}$ & $3.29\!\times\!10^{-3}$ & $2.32\!\times\!10^{-3}$ & 6 \\
\SPA{}-balanced & 1.61 & $1.83\!\times\!10^{-4}$ & $6.75\!\times\!10^{-7}$ & $1.83\!\times\!10^{-4}$ & $1.29\!\times\!10^{-4}$ & 6 \\
\SPA{}-accurate & 1.82 & $1.02\!\times\!10^{-5}$ & $6.68\!\times\!10^{-7}$ & $1.08\!\times\!10^{-5}$ & $7.19\!\times\!10^{-6}$ & 7 \\
Euston & 18.91 & $4.64\!\times\!10^{-4}$ & $1.66\!\times\!10^{-4}$ & $4.67\!\times\!10^{-4}$ & $3.46\!\times\!10^{-5}$ & 40 \\
\midrule
\multicolumn{7}{l}{\textit{QuickGELU, $R=20$}} \\
\SPA{}-fast & 4.44 & $8.40\!\times\!10^{-3}$ & $5.18\!\times\!10^{-6}$ & $8.41\!\times\!10^{-3}$ & $5.94\!\times\!10^{-3}$ & 7 \\
\SPA{}-balanced & 5.58 & $4.40\!\times\!10^{-4}$ & $4.80\!\times\!10^{-6}$ & $4.44\!\times\!10^{-4}$ & $3.11\!\times\!10^{-4}$ & 8 \\
\SPA{}-accurate & 6.44 & $2.31\!\times\!10^{-5}$ & $4.86\!\times\!10^{-6}$ & $2.74\!\times\!10^{-5}$ & $1.64\!\times\!10^{-5}$ & 8 \\
Euston & 21.61 & $1.87\!\times\!10^{-3}$ & $2.62\!\times\!10^{-3}$ & $2.92\!\times\!10^{-3}$ & $2.01\!\times\!10^{-4}$ & 42 \\
\midrule
\multicolumn{7}{l}{\textit{SiLU, $R=5$}} \\
\SPA{}-fast & 0.46 & $2.05\!\times\!10^{-2}$ & $7.51\!\times\!10^{-7}$ & $2.05\!\times\!10^{-2}$ & $1.44\!\times\!10^{-2}$ & 5 \\
\SPA{}-balanced & 0.69 & $1.78\!\times\!10^{-4}$ & $6.74\!\times\!10^{-7}$ & $1.79\!\times\!10^{-4}$ & $1.26\!\times\!10^{-4}$ & 6 \\
\SPA{}-accurate & 0.90 & $1.59\!\times\!10^{-6}$ & $6.71\!\times\!10^{-7}$ & $2.24\!\times\!10^{-6}$ & $1.13\!\times\!10^{-6}$ & 6 \\
Orion & 3.12 & $5.70\!\times\!10^{-11}$ & $1.22\!\times\!10^{-6}$ & $1.22\!\times\!10^{-6}$ & $6.61\!\times\!10^{-7}$ & 9 \\
ENSI & 1.15 & $2.09\!\times\!10^{-2}$ & $1.98\!\times\!10^{-6}$ & $2.09\!\times\!10^{-2}$ & $8.03\!\times\!10^{-3}$ & 7 \\
Sylph & 1.12 & $1.71\!\times\!10^{-4}$ & $9.00\!\times\!10^{-7}$ & $1.72\!\times\!10^{-4}$ & $1.23\!\times\!10^{-4}$ & 7 \\
THOR & 1.85 & $1.18\!\times\!10^{-7}$ & $3.57\!\times\!10^{-6}$ & $3.61\!\times\!10^{-6}$ & $1.40\!\times\!10^{-6}$ & 12 \\
\midrule
\multicolumn{7}{l}{\textit{SiLU, $R=20$}} \\
\SPA{}-fast & 1.13 & $8.94\!\times\!10^{-3}$ & $4.45\!\times\!10^{-6}$ & $8.94\!\times\!10^{-3}$ & $6.32\!\times\!10^{-3}$ & 7 \\
\SPA{}-balanced & 1.34 & $7.31\!\times\!10^{-4}$ & $3.95\!\times\!10^{-6}$ & $7.35\!\times\!10^{-4}$ & $5.17\!\times\!10^{-4}$ & 7 \\
\SPA{}-accurate & 1.46 & $5.99\!\times\!10^{-5}$ & $3.84\!\times\!10^{-6}$ & $6.37\!\times\!10^{-5}$ & $4.23\!\times\!10^{-5}$ & 8 \\
Orion & 3.12 & $1.74\!\times\!10^{-4}$ & $1.00\!\times\!10^{-5}$ & $1.67\!\times\!10^{-4}$ & $4.33\!\times\!10^{-5}$ & 9 \\
THOR & 1.84 & $3.31\!\times\!10^{-4}$ & $1.86\!\times\!10^{-5}$ & $3.40\!\times\!10^{-4}$ & $1.56\!\times\!10^{-4}$ & 12 \\
\midrule
\bottomrule
\end{longtable}
\normalsize

\section{Complete main-task results}
\label{app:fullresults}
The tables retain every reported main-task method and its failures. Gemma uses the current erf-target experiment, with Exact-erf and both original-target controls distinguished. Euston is not included in the Gemma-erf comparison. Llama data retain the original SiLU experiment. Raw source IDs are preserved in the accompanying files despite the paper-facing name change to \SPA{}.
\small\begin{longtable}{lrrrrr}\caption{All main-task MCQ comparisons. Gemma changes are relative to Exact-erf; Llama changes are relative to Native. Failed predictions count as incorrect.}\label{tab:mcqall}\\
\toprule Method & Acc. (\%) & $\Delta$ pp & 95\% CI & Disagr. (\%) & Invalid \\ \midrule\endfirsthead
\toprule Method & Acc. (\%) & $\Delta$ pp & 95\% CI & Disagr. (\%) & Invalid \\ \midrule\endhead
\multicolumn{6}{l}{\textit{Gemma-4-E2B-it: HellaSwag}} \\
Exact-erf & 57.81 & +0.00 & $[+0.0000,\,+0.0000]$ & 0.00 & 0 \\
Native-tanh & 57.76 & -0.05 & $[-0.2191,\,+0.1195]$ & 1.19 & 0 \\
Exact-tanh & 57.76 & -0.05 & $[-0.2191,\,+0.1195]$ & 1.19 & 0 \\
\SPA{}-fast & 57.90 & +0.09 & $[-0.1096,\,+0.2988]$ & 1.62 & 0 \\
\SPA{}-balanced & 57.91 & +0.10 & $[-0.0797,\,+0.2789]$ & 1.26 & 0 \\
\SPA{}-accurate & 57.73 & -0.08 & $[-0.2390,\,+0.0797]$ & 1.22 & 1 \\
NEXUS & 57.42 & -0.39 & $[-0.5975,\,-0.1792]$ & 1.77 & 57 \\
MOAI & 57.93 & +0.12 & $[-0.0698,\,+0.3187]$ & 1.43 & 0 \\
\midrule
\multicolumn{6}{l}{\textit{Gemma-4-E2B-it: MMLU}} \\
Exact-erf & 57.57 & +0.00 & $[+0.0000,\,+0.0000]$ & 0.00 & 0 \\
Native-tanh & 57.61 & +0.04 & $[-0.1282,\,+0.1995]$ & 1.51 & 0 \\
Exact-tanh & 57.61 & +0.04 & $[-0.1282,\,+0.1995]$ & 1.51 & 0 \\
\SPA{}-fast & 57.54 & -0.03 & $[-0.1995,\,+0.1354]$ & 1.63 & 0 \\
\SPA{}-balanced & 57.49 & -0.08 & $[-0.2422,\,+0.0855]$ & 1.48 & 0 \\
\SPA{}-accurate & 57.51 & -0.06 & $[-0.2279,\,+0.0926]$ & 1.50 & 0 \\
NEXUS & 51.69 & -5.88 & $[-6.3097,\,-5.4621]$ & 12.06 & 1500 \\
MOAI & 57.56 & -0.01 & $[-0.1995,\,+0.1638]$ & 1.76 & 0 \\
\midrule
\multicolumn{6}{l}{\textit{Gemma-4-E2B-it: Social IQa}} \\
Exact-erf & 48.67 & +0.00 & $[+0.0000,\,+0.0000]$ & 0.00 & 0 \\
Native-tanh & 48.46 & -0.20 & $[-0.5630,\,+0.1536]$ & 0.97 & 0 \\
Exact-tanh & 48.46 & -0.20 & $[-0.5630,\,+0.1536]$ & 0.97 & 0 \\
\SPA{}-fast & 48.57 & -0.10 & $[-0.4606,\,+0.2559]$ & 1.02 & 0 \\
\SPA{}-balanced & 48.52 & -0.15 & $[-0.5118,\,+0.1536]$ & 1.02 & 0 \\
\SPA{}-accurate & 48.62 & -0.05 & $[-0.5118,\,+0.4095]$ & 1.38 & 0 \\
NEXUS & 48.16 & -0.51 & $[-0.9724,\,-0.0511]$ & 1.94 & 17 \\
MOAI & 48.72 & +0.05 & $[-0.4095,\,+0.5118]$ & 1.43 & 0 \\
\midrule
\multicolumn{6}{l}{\textit{Gemma-4-E2B-it: RACE-high}} \\
Exact-erf & 44.23 & +0.00 & $[+0.0000,\,+0.0000]$ & 0.00 & 0 \\
Native-tanh & 44.00 & -0.23 & $[-0.5432,\,+0.0858]$ & 1.23 & 0 \\
Exact-tanh & 44.00 & -0.23 & $[-0.5432,\,+0.0858]$ & 1.23 & 0 \\
\SPA{}-fast & 44.23 & +0.00 & $[-0.4003,\,+0.4003]$ & 2.12 & 0 \\
\SPA{}-balanced & 44.03 & -0.20 & $[-0.5718,\,+0.1430]$ & 1.60 & 0 \\
\SPA{}-accurate & 44.08 & -0.14 & $[-0.4860,\,+0.2002]$ & 1.69 & 0 \\
NEXUS & 43.85 & -0.37 & $[-0.7719,\,+0.0286]$ & 2.54 & 35 \\
MOAI & 44.11 & -0.11 & $[-0.4860,\,+0.2573]$ & 1.74 & 0 \\
\midrule
\multicolumn{6}{l}{\textit{Llama-3.2-3B-Instruct: HellaSwag}} \\
Native & 71.65 & +0.00 & $[+0.0000,\,+0.0000]$ & 0.00 & 0 \\
Exact control & 71.65 & +0.00 & $[+0.0000,\,+0.0000]$ & 0.00 & 0 \\
\SPA{}-fast & 71.23 & -0.42 & $[-0.7867,\,-0.0398]$ & 4.52 & 0 \\
\SPA{}-balanced & 71.66 & +0.01 & $[-0.1295,\,+0.1494]$ & 0.64 & 0 \\
\SPA{}-accurate & 71.59 & -0.06 & $[-0.2092,\,+0.0797]$ & 0.69 & 0 \\
Orion & 71.65 & +0.00 & $[-0.1295,\,+0.1198]$ & 0.57 & 0 \\
ENSI & 70.44 & -1.20 & $[-1.5635,\,-0.8464]$ & 4.02 & 0 \\
Sylph & 71.67 & +0.02 & $[-0.1395,\,+0.1793]$ & 0.77 & 0 \\
THOR & 71.53 & -0.12 & $[-0.2490,\,+0.0100]$ & 0.56 & 0 \\
\midrule
\multicolumn{6}{l}{\textit{Llama-3.2-3B-Instruct: MMLU}} \\
Native & 62.16 & +0.00 & $[+0.0000,\,+0.0000]$ & 0.00 & 0 \\
Exact control & 62.16 & +0.00 & $[+0.0000,\,+0.0000]$ & 0.00 & 0 \\
\SPA{}-fast & 61.27 & -0.88 & $[-1.3603,\,-0.4130]$ & 11.63 & 0 \\
\SPA{}-balanced & 62.36 & +0.20 & $[+0.0284,\,+0.3632]$ & 1.63 & 0 \\
\SPA{}-accurate & 62.29 & +0.14 & $[-0.0428,\,+0.3063]$ & 1.65 & 0 \\
Orion & 62.09 & -0.06 & $[-0.2351,\,+0.1069]$ & 1.55 & 0 \\
ENSI & 61.79 & -0.37 & $[-0.7122,\,-0.0284]$ & 6.55 & 0 \\
Sylph & 62.23 & +0.07 & $[-0.1211,\,+0.2564]$ & 1.99 & 0 \\
THOR & 62.05 & -0.11 & $[-0.2920,\,+0.0713]$ & 1.83 & 28 \\
\midrule
\multicolumn{6}{l}{\textit{Llama-3.2-3B-Instruct: Social IQa}} \\
Native & 53.33 & +0.00 & $[+0.0000,\,+0.0000]$ & 0.00 & 0 \\
Exact control & 53.33 & +0.00 & $[+0.0000,\,+0.0000]$ & 0.00 & 0 \\
\SPA{}-fast & 51.43 & -1.89 & $[-2.8660,\,-0.9211]$ & 6.35 & 0 \\
\SPA{}-balanced & 53.07 & -0.26 & $[-0.6142,\,+0.0512]$ & 0.72 & 0 \\
\SPA{}-accurate & 53.33 & +0.00 & $[-0.4095,\,+0.4095]$ & 0.92 & 0 \\
Orion & 53.17 & -0.15 & $[-0.5118,\,+0.1536]$ & 0.72 & 0 \\
ENSI & 52.92 & -0.41 & $[-1.2283,\,+0.4095]$ & 4.40 & 0 \\
Sylph & 53.17 & -0.15 & $[-0.4606,\,+0.1536]$ & 0.77 & 0 \\
THOR & 53.07 & -0.26 & $[-0.6142,\,+0.0512]$ & 0.72 & 1 \\
\midrule
\multicolumn{6}{l}{\textit{Llama-3.2-3B-Instruct: RACE-high}} \\
Native & 49.60 & +0.00 & $[+0.0000,\,+0.0000]$ & 0.00 & 0 \\
Exact control & 49.60 & +0.00 & $[+0.0000,\,+0.0000]$ & 0.00 & 0 \\
\SPA{}-fast & 49.69 & +0.09 & $[-0.6576,\,+0.8291]$ & 7.69 & 0 \\
\SPA{}-balanced & 49.71 & +0.11 & $[-0.1430,\,+0.3717]$ & 1.00 & 0 \\
\SPA{}-accurate & 49.66 & +0.06 & $[-0.2002,\,+0.3145]$ & 0.86 & 0 \\
Orion & 49.66 & +0.06 & $[-0.2288,\,+0.3431]$ & 1.11 & 0 \\
ENSI & 49.51 & -0.09 & $[-0.6576,\,+0.4860]$ & 5.06 & 0 \\
Sylph & 49.80 & +0.20 & $[-0.0572,\,+0.4860]$ & 1.14 & 0 \\
THOR & 49.86 & +0.26 & $[+0.0000,\,+0.5432]$ & 1.14 & 0 \\
\midrule
\bottomrule
\end{longtable}
\normalsize
\footnotesize\begin{longtable}{lrrrrr}\caption{All main language-modeling comparisons. Gemma uses Exact-erf as reference. When windows fail, full-corpus PPL is omitted and paired differences refer only to shared valid windows.}\label{tab:lmall}\\
\toprule Method & PPL & $\Delta$NLL & 95\% CI & $\Delta$PPL (\%) & Lost \\ \midrule\endfirsthead
\toprule Method & PPL & $\Delta$NLL & 95\% CI & $\Delta$PPL (\%) & Lost \\ \midrule\endhead
\multicolumn{6}{l}{\textit{Gemma-4-E2B-it: WikiText-2}} \\
Exact-erf & 210.77 & $0$ & $[0,0]$ & +0.00 & 0 \\
Native-tanh & 210.59 & $-8.56\!\times\!10^{-4}$ & $[-1.4972,\,-0.2356]\!\times\!10^{-3}$ & -0.09 & 0 \\
Exact-tanh & 210.59 & $-8.56\!\times\!10^{-4}$ & $[-1.4972,\,-0.2356]\!\times\!10^{-3}$ & -0.09 & 0 \\
\SPA{}-fast & 203.95 & $-3.29\!\times\!10^{-2}$ & $[-3.3916,\,-3.1834]\!\times\!10^{-2}$ & -3.23 & 0 \\
\SPA{}-balanced & 211.13 & $1.71\!\times\!10^{-3}$ & $[+1.0625,\,+2.3713]\!\times\!10^{-3}$ & +0.17 & 0 \\
\SPA{}-accurate & 211.12 & $1.68\!\times\!10^{-3}$ & $[+0.9253,\,+2.4474]\!\times\!10^{-3}$ & +0.17 & 0 \\
NEXUS & -- & $-6.25\!\times\!10^{-3}$ & $[-7.1281,\,-5.3015]\!\times\!10^{-3}$ & -0.62 & 103 \\
MOAI & 216.83 & $2.84\!\times\!10^{-2}$ & $[+2.7491,\,+2.9292]\!\times\!10^{-2}$ & +2.88 & 0 \\
\midrule
\multicolumn{6}{l}{\textit{Gemma-4-E2B-it: C4-en}} \\
Exact-erf & 155.20 & $0$ & $[0,0]$ & +0.00 & 0 \\
Native-tanh & 154.95 & $-1.63\!\times\!10^{-3}$ & $[-1.9433,\,-1.3248]\!\times\!10^{-3}$ & -0.16 & 0 \\
Exact-tanh & 154.95 & $-1.63\!\times\!10^{-3}$ & $[-1.9433,\,-1.3248]\!\times\!10^{-3}$ & -0.16 & 0 \\
\SPA{}-fast & 150.89 & $-2.81\!\times\!10^{-2}$ & $[-2.8927,\,-2.7331]\!\times\!10^{-2}$ & -2.77 & 0 \\
\SPA{}-balanced & 155.52 & $2.07\!\times\!10^{-3}$ & $[+1.3246,\,+2.8708]\!\times\!10^{-3}$ & +0.21 & 0 \\
\SPA{}-accurate & 155.67 & $3.01\!\times\!10^{-3}$ & $[+2.2173,\,+3.8440]\!\times\!10^{-3}$ & +0.30 & 0 \\
NEXUS & -- & $-4.78\!\times\!10^{-3}$ & $[-5.8352,\,-3.6385]\!\times\!10^{-3}$ & -0.48 & 534 \\
MOAI & 159.03 & $2.44\!\times\!10^{-2}$ & $[+2.3770,\,+2.5012]\!\times\!10^{-2}$ & +2.47 & 0 \\
\midrule
\multicolumn{6}{l}{\textit{Llama-3.2-3B-Instruct: WikiText-2}} \\
Native & 12.17 & $0$ & $[0,0]$ & +0.00 & 0 \\
Exact control & 12.17 & $0$ & $[0,0]$ & +0.00 & 0 \\
\SPA{}-fast & 12.63 & $3.64\!\times\!10^{-2}$ & $[+3.4735,\,+3.7978]\!\times\!10^{-2}$ & +3.70 & 0 \\
\SPA{}-balanced & 12.18 & $2.42\!\times\!10^{-4}$ & $[+0.5113,\,+4.8321]\!\times\!10^{-4}$ & +0.02 & 0 \\
\SPA{}-accurate & 12.18 & $1.50\!\times\!10^{-4}$ & $[-0.2761,\,+3.8233]\!\times\!10^{-4}$ & +0.02 & 0 \\
Orion & 12.17 & $5.50\!\times\!10^{-5}$ & $[-0.6906,\,+1.8306]\!\times\!10^{-4}$ & +0.01 & 0 \\
ENSI & 12.93 & $6.04\!\times\!10^{-2}$ & $[+5.8842,\,+6.1914]\!\times\!10^{-2}$ & +6.22 & 0 \\
Sylph & 12.20 & $2.10\!\times\!10^{-3}$ & $[+1.7393,\,+2.4959]\!\times\!10^{-3}$ & +0.21 & 0 \\
THOR & -- & $-4.18\!\times\!10^{-5}$ & $[-1.7114,\,+0.8497]\!\times\!10^{-4}$ & -0.00 & 1 \\
\midrule
\multicolumn{6}{l}{\textit{Llama-3.2-3B-Instruct: C4-en}} \\
Native & 18.38 & $0$ & $[0,0]$ & +0.00 & 0 \\
Exact control & 18.38 & $0$ & $[0,0]$ & +0.00 & 0 \\
\SPA{}-fast & 18.90 & $2.77\!\times\!10^{-2}$ & $[+2.6260,\,+2.9258]\!\times\!10^{-2}$ & +2.81 & 0 \\
\SPA{}-balanced & 18.39 & $3.43\!\times\!10^{-4}$ & $[+0.9648,\,+6.6571]\!\times\!10^{-4}$ & +0.03 & 0 \\
\SPA{}-accurate & 18.39 & $8.49\!\times\!10^{-4}$ & $[+0.2652,\,+1.5919]\!\times\!10^{-3}$ & +0.08 & 0 \\
Orion & 18.38 & $7.50\!\times\!10^{-5}$ & $[+0.0324,\,+1.4646]\!\times\!10^{-4}$ & +0.01 & 0 \\
ENSI & 19.46 & $5.73\!\times\!10^{-2}$ & $[+5.5923,\,+5.8805]\!\times\!10^{-2}$ & +5.90 & 0 \\
Sylph & 18.45 & $3.73\!\times\!10^{-3}$ & $[+2.9644,\,+4.6708]\!\times\!10^{-3}$ & +0.37 & 0 \\
THOR & -- & $3.63\!\times\!10^{-5}$ & $[-0.6217,\,+1.5697]\!\times\!10^{-4}$ & +0.00 & 6 \\
\midrule
\bottomrule
\end{longtable}
\normalsize

\section{Precision diagnostics and compact summaries}
\label{app:precision}
Coefficient rounding changes the frozen residual coefficients at one through five decimal places. CKKS precision instead keeps the full coefficients and varies scale/interior-prime bit sizes. The balanced R20 degrees are $k=31$ for erf-GELU, $k=39$ for QuickGELU, and $k=23$ for SiLU. Rounding may make terms zero, and the evaluator can prune them; changes in runtime therefore also reflect pruning, not merely the number of decimal places retained.

\begin{table}[ht]
\centering\footnotesize\setlength{\tabcolsep}{3pt}
\caption{Balanced language-modeling results. Each comparison is against its own model/target reference. NLL changes are token-weighted; confidence-interval endpoints use four decimal places with a common power-of-ten factor.}
\label{tab:lmbal}\begin{tabular}{llrrrrr}
\toprule
Model & Corpus & Ref. PPL & SPA PPL & $\Delta$NLL & 95\% CI & $\Delta$PPL (\%) \\
\midrule
Gemma-erf & WikiText-2 & 210.77 & 211.13 & $1.71\!\times\!10^{-3}$ & $[+1.0625,\,+2.3713]\!\times\!10^{-3}$ & +0.17 \\
Gemma-erf & C4-en & 155.20 & 155.52 & $2.07\!\times\!10^{-3}$ & $[+1.3246,\,+2.8708]\!\times\!10^{-3}$ & +0.21 \\
Llama & WikiText-2 & 12.17 & 12.18 & $2.42\!\times\!10^{-4}$ & $[+0.5113,\,+4.8321]\!\times\!10^{-4}$ & +0.02 \\
Llama & C4-en & 18.38 & 18.39 & $3.43\!\times\!10^{-4}$ & $[+0.9648,\,+6.6571]\!\times\!10^{-4}$ & +0.03 \\
\bottomrule
\end{tabular}

\end{table}
\begin{table}[ht]
\centering\small\setlength{\tabcolsep}{4pt}
\caption{Selected paired MCQ diagnostics, relative to Exact-erf for Gemma and Native for Llama. Prediction disagreement differs from the change in correctness.}
\label{tab:paired}\begin{tabular}{llrrrr}
\toprule
Model & Method & $\Delta$ pp & 95\% CI & Disagr. (\%) & Invalid \\
\midrule
Gemma-erf & \SPA{}-balanced & -0.04 & $[-0.1422,\,+0.0711]$ & 1.39 & 0 \\
Gemma-erf & MOAI & +0.02 & $[-0.0948,\,+0.1422]$ & 1.63 & 0 \\
Gemma-erf & NEXUS & -3.01 & $[-3.2334,\,-2.7898]$ & 6.77 & 1609 \\
Llama & \SPA{}-balanced & +0.09 & $[-0.0068,\,+0.1964]$ & 1.16 & 0 \\
\bottomrule
\end{tabular}

\end{table}
\begin{table}[ht]
\centering\small\setlength{\tabcolsep}{4pt}
\caption{Archived Llama-SiLU coefficient-rounding results. Accuracy changes are pooled percentage points; NLL changes use C4-en.}
\label{tab:round}\begin{tabular}{rrrr}
\toprule
Decimals & Llama $\Delta$Acc. (pp) & C4 $\Delta$NLL & MCQ invalid \\
\midrule
1 & -0.99 & $5.50\!\times\!10^{-2}$ & 0 \\
2 & -0.44 & $1.16\!\times\!10^{-2}$ & 0 \\
3 & +0.07 & $7.02\!\times\!10^{-4}$ & 0 \\
4 & +0.02 & $3.45\!\times\!10^{-4}$ & 0 \\
5 & +0.06 & $3.78\!\times\!10^{-4}$ & 0 \\
\bottomrule
\end{tabular}

\end{table}

\section{Experimental environment and code availability}
\label{app:hardware}
\begin{center}\small
\begin{tabularx}{\linewidth}{@{}lX@{}}
\toprule
Item & Specification \\
\midrule
CPU (CKKS kernels) & Intel Xeon Gold 6530 \\
GPU (plaintext models) & NVIDIA RTX PRO 6000 Blackwell Workstation Edition \\
Homomorphic library & SEAL 4.1 \\
Code repository & \CodeRepositoryLink \\
\bottomrule
\end{tabularx}
\end{center}

\section{Coefficient construction and evaluation algorithms}
\label{app:algorithms}
\subsection{Recorded numerical construction}
For each target, radius, and candidate $k$, Equation~\eqref{eq:discretefit} is solved using SciPy \texttt{linprog}\allowbreak\texttt{(method="highs")} in float64 with library-default tolerances. All $k+1$ coefficients are free. This is a finite-node minimax problem, not a least-squares problem.

During refinement of the ciphertext coefficients, the error is evaluated on a uniform 300,001-point grid in $t$. Endpoints and local extrema satisfying $|e|\ge0.80\max|e|$, together with a maximum-error point, are candidates for addition. Refinement stops if no new nodes are added, if the sampled maximum is at most $(1+2\times10^{-8})\epsilon_{\rm LP}$, or if the iteration cap is reached. A separate 300,001-point grid in the original input interval checks the activation error.

The degree search fixes $m=4$ for R5 and $m=8$ for R20/wide, then increases the block count $B=1,2,\ldots$ with candidate degree $k=Bm-1$. Thus $B=\lceil(k+1)/m\rceil$, matching Section~\ref{sec:psconstruction}; the lowercase $b$ in that section is only a block index.

\subsection{Chebyshev division and the PS blocks}
Represent $F=\sum_i f_iT_i$ and divide by $T_m$. Initialize the remainder coefficients as $r\gets f$ and set the quotient to zero. Process indices $i$ in descending order from the degree of $F$ to $m$. At $i=m$, transfer $r_m$ to the constant quotient coefficient and set $r_m=0$. At $i>m$, with $a=r_i$, use
\[
 aT_i(t)=\bigl(2aT_{i-m}(t)\bigr)T_m(t)-aT_{|i-2m|}(t).
\]
Thus add $2a$ to quotient coefficient $i-m$, set the old coefficient $r_i$ to zero, and subtract $a$ from remainder coefficient $|i-2m|$. The latter index is strictly smaller than $i$, so the procedure terminates. It returns $F=T_mQ+B$ with $\deg B<m$. Repeated application gives~\eqref{eq:chebdivision} and~\eqref{eq:ps} for any input coefficient vector. This offline conversion is polynomial division, not direct grouping of Chebyshev coefficients.

For $k=7,m=4$, let $Q_c=\sum_{j=0}^{7}c_jT_j$. The two blocks are
\begin{align*}
 B_0&=c_0+(c_1-c_7)T_1+(c_2-c_6)T_2+(c_3-c_5)T_3,\\
 B_1&=c_4+2c_5T_1+2c_6T_2+2c_7T_3,
 \qquad Q_c=B_0+T_4B_1.
\end{align*}
The required Chebyshev values are computed using
\[
 T_2=2t^2-1,\qquad T_3=2tT_2-t,\qquad T_4=2T_2^2-1.
\]
Including the input square $x^2$ and the final product $T_4B_1$ yields five CT$\times$CT operations. Normalization, the retained $x/2$ term, and multiplication by the six nonconstant block coefficients require eight PT$\times$CT operations in the recorded dense schedule. Its six consumed levels include two for input formation, two for the longest basis path, one for coefficient products, and one for final recombination. Counts are for this schedule and need not apply after pruning or a different scale-management strategy.

\end{document}